\documentclass[11pt]{article}

\usepackage[final]{acl}

\usepackage{times}
\usepackage{latexsym}

\usepackage[T1]{fontenc}

\usepackage[utf8]{inputenc}

\usepackage{microtype}

\usepackage{inconsolata}

\usepackage{graphicx}

\usepackage{amsmath,amsfonts}
\usepackage{pgf}
\usepackage[table, dvipsnames]{xcolor}  
\usepackage{colortbl}  
\usepackage{subcaption}
\usepackage{mathtools}
\usepackage{xspace}
\usepackage{array}
\usepackage{ragged2e}
\usepackage{longtable}
\usepackage{algorithm}  
\usepackage{algpseudocode}  
\usepackage{tikz}
\usetikzlibrary{fit,arrows,shapes,automata,backgrounds,petri}
\usetikzlibrary{arrows.meta}
\usepackage{multirow}
\usepackage{rotating}  
\usepackage{makecell}  
\usepackage{listings}
\usepackage{caption}
\usepackage{fancyvrb}
\usepackage{bbding}
\usepackage{pifont}
\usepackage{wasysym}
\usepackage{tabularx}
\usepackage{pdfpages}
\usepackage{colortbl}
\usepackage{enumitem}
\usepackage{pifont}
\usepackage{makecell}
\usepackage{comment}
\usepackage{booktabs}
\usepackage{textcomp}
\usepackage{lscape}
\usepackage{verbatim}
\usepackage{textcomp}
\usepackage[most]{tcolorbox}
\usepackage{hyperref}
\usepackage{soul}
\usepackage{adjustbox}
\usepackage{url}
\usepackage{float}

\newcommand{\system}{\text{\fontfamily{lmtt}\selectfont LogicEval}\xspace}
\newcommand{\dataset}{\text{\fontfamily{lmtt}\selectfont LogicDS}\xspace}

\newcommand{\tianchang}[1]{\textcolor{Orchid}{[Tianchang: #1]}}

\definecolor{PromptGreen}{RGB}{0,140,120}
\definecolor{DotsGray}{RGB}{90,90,90}

\newenvironment{promptblock}
{%
  \par\noindent
  \begingroup
  \ttfamily\small
  \color{PromptGreen}
  \setlength{\parindent}{0pt}
  \setlength{\parskip}{2pt}
}
{%
  \endgroup
  \par
}

\title{LogicEval: A Systematic Framework for Evaluating  Automated Repair Techniques for Logical Vulnerabilities in Real-World Software}

\author{
    Syed Md Mukit Rashid,
    Abdullah Al Ishtiaq,
    Kai Tu,
    Yilu Dong,
    Tianwei Wu,\\
    \textbf{Ali Ranjbar},
    \textbf{Tianchang Yang},
    \textbf{Najrin Sultana},
    \textbf{Shagufta Mehnaz},
    \textbf{Syed Rafiul Hussain}\\
    The Pennsylvania State University\\
    \texttt{\{szr5848, abdullah.ishtiaq, kjt5562, yiludong, tvw5452,}\\
    \texttt{aranjbar, tzy5088, nks5814, smehnaz, hussain1\}@psu.edu}
}

\definecolor{codecomment}{rgb}{0.57,0.63,0.63}
\definecolor{codegray}{rgb}{0.5,0.5,0.5}
\definecolor{codestring}{rgb}{0.16,0.63,0.59}
\definecolor{backcolour}{rgb}{0.98,0.98,0.96}
\definecolor{codeemph}{rgb}{0.34,0.43,0.46}
\definecolor{codetype}{rgb}{0.79,0.29,0.086}
\definecolor{codefunction}{rgb}{0.15,0.54,0.82}
\definecolor{codekeyword}{rgb}{0.52,0.6,0.0}

\lstdefinestyle{logiceval}{
    classoffset=0,
    otherkeywords={chacha20_poly1305_ctrl,ssl3_accept,ssl3_read_bytes},
    morekeywords={chacha20_poly1305_ctrl,ssl3_accept,ssl3_read_bytes},
    keywordstyle=\color{codefunction},
    classoffset=1,
    backgroundcolor=\color{backcolour},
    commentstyle=\itshape\color{codecomment},
    keywordstyle=\color{codekeyword},
    numberstyle=\tiny\color{codegray},
    stringstyle=\color{codestring},
    basicstyle=\ttfamily\scriptsize,
    breakatwhitespace=false,
    breaklines=true,
    captionpos=b,
    keepspaces=true,
    numbers=left,
    numbersep=5pt,
    xleftmargin=10pt,
    linewidth=\linewidth,
    showspaces=false,
    showstringspaces=false,
    showtabs=false,
    tabsize=2,
    emph={int,char,double,float,unsigned,void,bool,uint32_t,uint8_t,uint16_t,ushort,byte,uint,typedef,struct,static},
    emphstyle=\bfseries\color{codeemph},
    postbreak=\mbox{\textcolor{codegray}{$\hookrightarrow$}\space},
}
\newcommand{\lstbg}[3][0pt]{%
    {\fboxsep=#1%
        \colorbox{#2}{\makebox[\dimexpr\linewidth-2\fboxsep][l]{\strut #3}}%
    }%
}
\newcommand{\lstbgt}[3][0pt]{%
    {\fboxsep=#1%
        \colorbox{#2}{\strut #3}%
    }%
}
\newcommand{\hashchar}{\char`\#}
\newcommand{\opencb}{\char`\{}
\newcommand{\closecb}{\char`\}}

\newcommand{\savelinenumbers}{%
    \let\origthelstnumber\thelstnumber
}
\def\createlinenumber#1#2{
    \edef\thelstnumber{%
        \unexpanded{%
            \ifnum#1=\value{lstnumber}\relax
              #2%
            \fi}%
        \ifx\thelstnumber\relax\else
        \expandafter\unexpanded\expandafter{\thelstnumber}%
        \fi
    }
}
\newcommand{\resetlinenumbers}{%
    \let\thelstnumber\origthelstnumber
}

\lstdefinelanguage{diff}{
    breakatwhitespace=false,
    basicstyle=\ttfamily\scriptsize,
    breaklines=true,
    captionpos=b,
    keepspaces=true,
    numbers=left,
    numbersep=5pt,
    showspaces=false,
    showstringspaces=false,
    showtabs=false,
    tabsize=2,
    xleftmargin=10pt,
    linewidth=\linewidth,
    morecomment=[f][\lstbg{red!20}]-,
    morecomment=[f][\lstbg{green!20}]+,
    morecomment=[f][\textit]{@@},
}

\begin{document}

\maketitle
\begin{abstract}
Logical vulnerabilities in software stem from flaws in program logic rather than memory safety, which can lead to critical security failures. Although existing automated program‐repair techniques primarily focus on repairing memory‐corruption vulnerabilities, they struggle with logical vulnerabilities because of their limited semantic understanding of the vulnerable code and its expected behavior. On the other hand, recent successes of large language models (LLMs) in understanding and repairing code are promising. However, no framework currently exists to analyze the capabilities and limitations of such techniques for logical vulnerabilities. We aim to systematically evaluate both traditional and LLM-based repair approaches for addressing real-world logical vulnerabilities. To facilitate our assessment, we created the first-ever dataset, LogicDS, comprising 122 logical vulnerabilities that reflect tangible security impact. We also developed a systematic framework, LogicEval, to evaluate patches for logical vulnerabilities. Evaluations suggest that compilation and testing failures are primarily driven by prompt sensitivity, loss of code context, and difficulty in patch localization. \footnote{To appear in ACL 2026 Main Conference.}
\end{abstract}

\section{Introduction}\label{introduction}

Logical vulnerabilities in software pose significant risks because they stem from incorrect implementation of the program's logic/functionality, rather than violations of language safety measures, such as memory corruptions. These vulnerabilities can be exploited to induce critical security and privacy implications, including authentication bypass \cite{tu2024logic}, sensitive data leakage \cite{ranjbar_loris_2025}, or system operations disruptions \cite{corecrisis}, often without triggering traditional security defenses, e.g., address sanitizers~\cite{song2019sok}. Several works \cite{felmetsger2010toward, deepa2018detlogic} have focused on identifying logical vulnerabilities in complex software systems and evaluating LLM performance to identify them \cite{ullah2023can}. However, little attention has been given to automatically repairing them, leaving vulnerable software exposed to significant risk.

Automatically repairing logical vulnerabilities presents unique challenges compared to fixing memory-corruption vulnerabilities. First, logical vulnerabilities do not follow consistent, reusable repair templates/patterns~\cite{li2025sok}. Generating a correct patch often requires deep semantic understanding of the vulnerable code, its surrounding context, and the compromised functionality or property. They also do not necessarily lead to crashes or illegal memory access, thus conventional signals such as compilation logs, runtime logs, or memory sanitizers~\cite{song2019sok} provide limited help for localization. Finally, generating effective test cases to assess their patch correctness is difficult because these traditional tools offer little guidance.

Since manual vulnerability repair is labor-intensive and time-consuming \cite{li2017large}, prior work has proposed automated techniques for generating patches \cite{gao2021beyond, jiang2021cure, le2012systematic}. However, most of these approaches target memory-safety violations, making them ineffective or unreliable for addressing logical vulnerabilities. Template-based~\cite{lin2007autopag, novark2007exterminator, cheng2019automatic} and search-based~\cite{le2012systematic, van2018static} patch generation techniques rely on recognizable repair patterns, which limit generalization to logical vulnerabilities whose patches are diverse and context-specific. Deep learning-based methods \cite{chen2019sequencer, jiang2021cure} seem a plausible option, but their effectiveness degrades on previously unseen patterns.

Recently, Large Language Models (LLMs) trained on vast training data have demonstrated the ability to generate high-quality outputs across tasks such as text summarization \cite{jin2024comprehensive, urlana2024controllable} and question answering \cite{saito2024unsupervised, yixing2024chain}. Due to similarities between code and natural language, researchers have explored LLMs for code generation \cite{ni2025tree, liu2024your}, analysis \cite{fang2024large}, and repair \cite{jin2023inferfix, kulsum2024case, wang2024intervenor}. LLMs can capture program syntax and semantics and reason about program behaviors. Recently, LLM-based approaches have attempted to localize vulnerabilities \cite{li2024attention}, generate repair patches \cite{jin2023inferfix, kulsum2024case, pearce2023examining}, and assess patch correctness \cite{zhou2024leveraging}, underscoring LLMs' potential to repair logical vulnerabilities.

However, to our knowledge, there is no systematic framework for analyzing the capabilities and limitations of automated vulnerability repair (AVR) for logical vulnerabilities. A targeted evaluation is critical for advancing AVR beyond memory-safety issues and toward the more subtle domain of logical vulnerabilities. To this end, we develop a systematic framework for evaluating repair techniques for logical vulnerabilities. We examine the performance of state-of-the-art non-learning-based, learning-based, and off-the-shelf LLM approaches, and study how prompting strategies, repair complexity, and auxiliary vulnerability information influence repair outcomes. We also identify the key challenges LLM-based approaches must address to improve performance.

Unfortunately, existing vulnerability datasets~\cite{just2014defects4j, gao2021beyond, fan2020ac, jimenez2023swe} primarily focus on memory-safety bugs and lack representative samples of real-world logical vulnerabilities, thereby limiting their usefulness for our purposes. To address this gap, we first construct \dataset, the first-ever curated dataset containing 61 real-world logical vulnerabilities including vulnerable and fixed code, vulnerability descriptions, and, when available, behavioral specifications and developer-authored repair rationales.

Building on this dataset, we then design \system, an end-to-end evaluation framework that takes localized vulnerable code and auxiliary inputs (e.g., context, specifications, repair descriptions) and evaluates AVR techniques, including both traditional tools and LLM-based methods, on their ability to synthesize correct patches. \system supports diverse prompt configurations and enforces structured patch outputs for seamless grafting into source code. It evaluates patches using compilation and testing pipelines and introduces automated reasoning to assess patch quality. Together, these capabilities enable \system to systematically characterize traditional and LLM-based AVR approaches, as well as off-the-shelf LLMs, for patch generation and assessment on logical vulnerabilities. Through extensive empirical analysis, we identify key strengths, limitations, and failure modes of LLMs and offer insights to guide future AVR development for logical vulnerabilities.


\section{Logical Vulnerabilities}\label{sec:logical_vulnerabilities}

A logical vulnerability $\vartheta$ is a logical flaw 
in a program $P$ that leads the program to deviate from an expected behavior $E$ and compromises its security in terms of confidentiality, integrity, or availability.

\begin{lstlisting}[
    language=C,
    caption={Fix for CVE-2019-1543~\cite{cve20191543_vul}. Lines prefixed with \texttt{++} denote additions, whereas \texttt{--} indicate deletions.},
    label=lst:example1,
    escapeinside={(*@}{@*)},
    float=ht,
    aboveskip=0cm,
    belowskip=-0.5cm
]
typedef struct {
...
(*@\lstbg{green!20}{++ \hashchar define CHACHA20\_POLY1305\_MAX\_IVLEN 12}@*)
...

static int chacha20_poly1305_ctrl(EVP_CIPHER_CTX *ctx, int type, int arg)
...
     return 1;

   case EVP_CTRL_AEAD_SET_IVLEN:
(*@\lstbg{red!20}{--~~~if (arg <= 0 || arg > CHACHA\_CTR\_SIZE)}@*)
(*@\lstbg{green!20}{++~~~if (arg <= 0 || arg > CHACHA20\_POLY1305\_MAX\_IVLEN)}@*)
       return 0;
     actx->nonce_len = arg;
     return 1;
...
\end{lstlisting}

Listing~\ref{lst:example1} shows a real-world logical vulnerability in OpenSSL’s \texttt{chacha20\_poly1305\_ctrl}, which handles ChaCha20-Poly1305 control operations. RFC 7539~\cite{rfc_7539} requires a unique 96-bit (12-byte) nonce. However, OpenSSL allows variable-length nonces, pads those shorter than 12 bytes with leading zeros, and even accepts lengths up to 16 bytes, using only the last 12 bytes and silently discarding the rest. This improper validation of the \texttt{arg} parameter risks nonce reuse and potentially enables severe cryptographic attacks~\cite{cve20191543_vul}. 

The expected behavior of logical vulnerabilities is often described by design specifications (e.g., RFCs) or implementation documents. However, they can also stem from implicit knowledge (e.g., privilege escalation in CVE-2024-25420 \cite{cve202425420}). Logical vulnerabilities may compromise an implementation’s security assumptions even when the program is free from memory-safety issues. Logical vulnerabilities usually stem from incorrect or missing control flow or validation logic (e.g., skipping required state transitions, omitting essential preconditions, placing checks in the wrong location, or accepting values outside the intended range). They do not exhibit consistent fix patterns. As such, resolving them requires a deep understanding of the program logic and its intended behavior.

\section{Related Works}
\noindent\textbf{AVR approaches.} Automatic Vulnerability Repair (AVR) techniques aim to propose a patch that (i) eliminates an identified vulnerability and (ii) preserves the original functionality after the fix. Existing AVR approaches can be broadly classified into \emph{non-learning-based}, \emph{learning-based}, and \emph{LLM-based} approaches. Among non-learning-based AVR approaches, template-guided approaches \cite{shaw2014automatically,huang2019using,xing2024if,zhang2022example} rely on patterns of vulnerability properties or historical patches to generate patches, and thus usually excel where known fix templates can be reliably applied. In contrast, fixes for logical vulnerabilities typically lack such common patterns. Constraint-based approaches extract program constraints through static analysis \cite{oh2018memfix,chida2022repairing}, symbolic execution \cite{shariffdeen2021concolic}, or dynamic analysis \cite{xuan2016nopol,agrawal1990dynamic} against a set of test suites. However, each logical vulnerability typically has its own distinct expected behavior that does not necessarily map to the program structure and often provides only limited exploit traces. Search-based approaches \cite{le2012systematic,marginean2019sapfix,jiang2018shaping,le2016history} explore a hypothetical search space for repair through mutations in existing code or by extracting similar code from other functions or source files within a project; such mutations perform poorly for real-world logical vulnerabilities. Deep learning-based AVR approaches \cite{chen2019sequencer,jiang2021cure,jiang2023knod} treat program repair as a neural machine translation task. Although these approaches perform better than non-learning-based approaches, they require a dataset of vulnerabilities and their fixes for training and attempt to extract recurring repair patterns from source code. Recently, several LLM-based approaches have been proposed to generate repair patches \cite{jin2023inferfix,kulsum2024case,pearce2023examining}. LLMs appear promising for repairing logical vulnerabilities because they can understand security invariants, though they face limitations when evaluating complex constraints. However, to our knowledge, no prior research has systematically analyzed their capabilities and limitations in this specific context. Further details regarding existing AVR techniques and their limitations are provided in Appendix \ref{sec:appendix:characterizing_avr}.

\noindent\textbf{AVR evaluation frameworks.} Existing evaluations on LLMs for security vulnerabilities assess the security of generated code \cite{pearce2025asleep}, study LLM assistance in writing code \cite{sandoval2023lost}, or evaluate LLMs for vulnerability identification and related security tasks \cite{deng2024pentestgpt,ullah2023can}. Prior work also provides taxonomies and benchmarks for AVR approaches \cite{li2025sok}, but does not address logical vulnerabilities or the limitations of existing AVR approaches for them. Most related to our work is that of Pearce et al.~\cite{pearce2023examining}, which evaluates the performance of LLMs towards repairing security vulnerabilities with zero-shot prompts. However, their evaluation framework does not take auxiliary vulnerability information as input, lacks automated metrics to identify reasonable patches, and uses address sanitizers and CodeQL to test patches, which are not suitable for evaluating logical vulnerabilities.

\section{\system{}: Evaluation Framework}\label{sec:methodology}

\begin{figure}
    \centering
    \includegraphics[width=\linewidth]{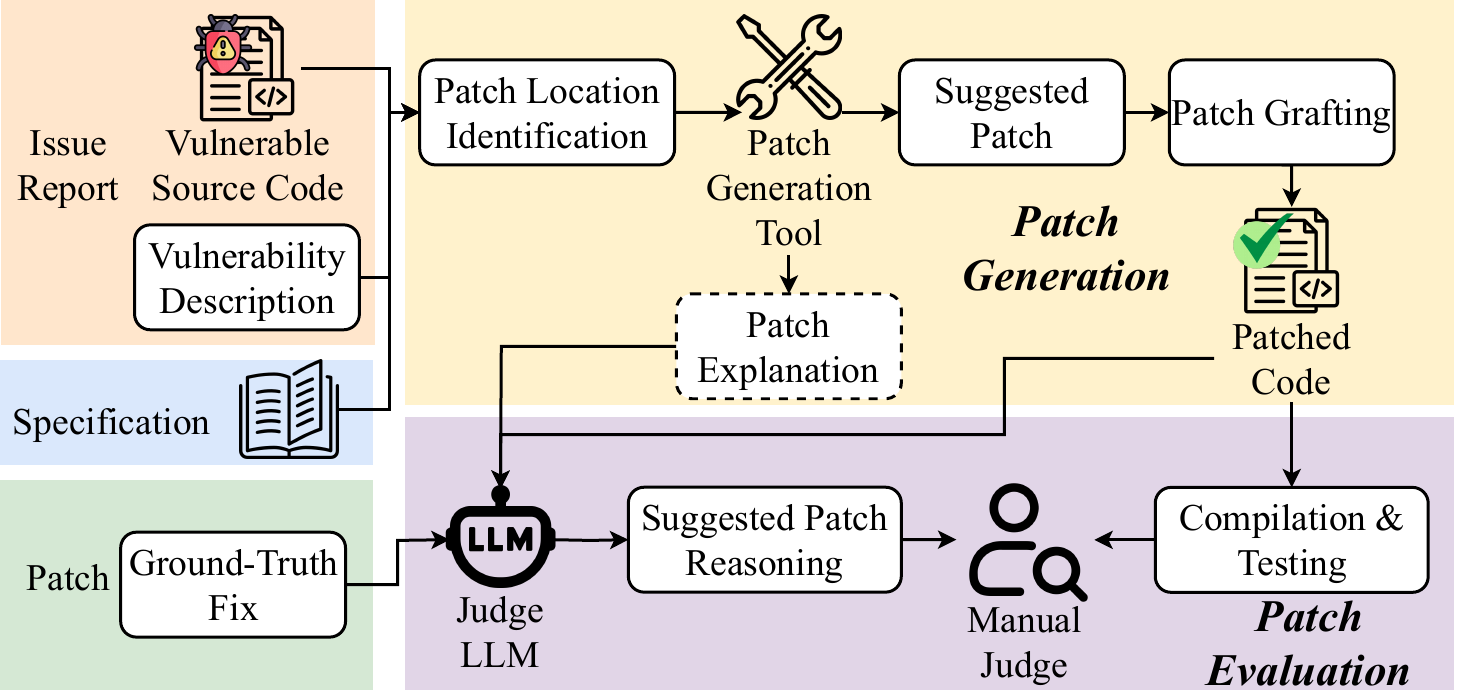}
    \caption{Overview of \system 
    }
    \label{fig:pipeline_evaluation}
    \vspace{-5mm}
\end{figure}

To address limitations of existing evaluation approaches, we develop \system, a systematic end-to-end framework that automatically generates and evaluates patches for logical vulnerabilities. We also define metrics to assess patch correctness and quality. Using the generated patches and evaluation results, \system enables analysis of the capabilities and limitations of different AVR approaches. Figure~\ref{fig:pipeline_evaluation} outlines the \system workflow.

\subsection{Inputs for \system}
\label{subsubsec:inputs_patch_generation}
We assume the following inputs to \system:

\noindent\textbf{\textit{Vulnerable ($S$) and Fixed ($F$) Source Code.}} We provide code from the vulnerable implementation source code $S$ as input to the repair framework under evaluation. After patch generation, we assess patch quality by comparing generated patches against a ground-truth patch $F$.

\noindent\textbf{\textit{Vulnerability Description ($D$).}}We assume a natural text description of the vulnerability is available to clarify how it leads to deviations from expected behavior and violates security guarantees. 

\noindent\textbf{\textit{Behavioral Specification ($V_S$).}} In some experimental settings, we additionally provide natural-text behavioral specifications (e.g., extracted from RFCs or formal implementation documentation). Since such specifications are not always available for all logical vulnerabilities, this input is optional and is only provided in certain experiments.

\noindent\textbf{\textit{Context ($V_{ctx}$).}} We define context $V_{ctx}$ as local variable declarations at the start of the vulnerable function and global/static definitions at the top of the file/class. We optionally provide $V_{ctx}$ in some experiments, as it may help repair techniques generate more plausible fixes.

\noindent\textbf{\textit{Compilation ($C$) and Testing ($T$) Scripts.}} We provide a script to compile the full project with any suggested patch. If available, we also provide a testing script $T$ that runs existing testcases against the project, often including a proof-of-concept (PoC) exploit to check whether the vulnerability persists.

\subsection{Patch Localization} \label{subsubsec:vulnerability_localization}
To generate a patch for a logical vulnerability, a developer must first \emph{locate} the code segment that requires modification (i.e., patch localization). Most existing AVR approaches~\cite{jiang2023knod, kulsum2024case, jiang2021cure, lutellier2020coconut} either assume perfect localization or rely on automated fault localization. Prior work on fault localization~\cite{campos2012gzoltar, jones2005empirical, oichai_fl} mainly targets general bugs or memory-safety issues and is not designed for logical vulnerabilities, which are harder to localize, as automated techniques (e.g., address sanitizers) cannot identify code that triggers them. Moreover, no existing method achieves perfect accuracy, and localization errors frequently lead to incorrect or ineffective AVR patches~\cite{jiang2018shaping, jiang2023knod}. Furthermore, both existing AVR techniques and off-the-shelf LLMs (\S\ref{subsec:llm_performance}) already struggle with logical vulnerabilities. To isolate patch-generation performance, we restrict our evaluation to single-hunk fixes within a single function and assume \emph{perfect patch localization} by manually identifying the precise vulnerable region.

However, real-world patches for security-relevant logical vulnerabilities often span multiple locations. We observe that most such patches contain a \emph{core fix}: a contiguous block of statements that implements the essential logic to eliminate the vulnerability, while other changes (e.g., variable declarations, helper functions, or supporting assignments) primarily support this core fix (examples provided in Appendix \ref{sec:appendix:example_codes}). Hence, we treat repairing the core fix as addressing a single-hunk vulnerability, motivated by (i) evidence that AVR performs better on single-hunk patches~\cite{li2025sok} and (ii) the intuition that multi-hunk vulnerabilities can often be handled by generating patches independently at each location. We provide additional examples and our identification procedure in the Appendix \ref{sec:appendix:locating_core_fix}.  

\subsection{Patch Generation for \system}\label{subsubsec:patch_generation_grafting}
\system leverages the provided inputs and localized vulnerability to generate patches using the AVR technique under evaluation $M$. For baseline approaches (Table~\ref{tab:repair_approaches_summary}), we provide vulnerable source code $V$ and the required localization (either the vulnerable block $V_b$ or function $V_s$). When evaluating off-the-shelf LLMs, we construct prompts along the following dimensions using the inputs in \S\ref{subsubsec:inputs_patch_generation}:

\noindent\textbf{\textit{Portion of source code.}} We provide either the vulnerable block $V_b$ or vulnerable function $V_f$; in some experiments, we also include context $V_{ctx}$.

\noindent\textbf{\textit{Auxiliary information.}} To assist LLM-based repair, we provide one or more of the vulnerability description $D$, behavioral specification $S$, and repair description $R$. We vary these inputs across experiments to measure how auxiliary information affects patch correctness and quality.

\noindent\textbf{\textit{LLM Configuration.}} We evaluate prompting strategies, including zero-shot, few-shot, and chain-of-thought. We also configure temperature and explore task-oriented and role-oriented prompting styles.

\noindent\textbf{\textit{Patch grafting.}} While constructing prompts, we instruct the LLM to place the suggested repair within tags and to output code that can directly replace the provided vulnerable function or block. After receiving the output, we extract the tagged code and replace the annotated vulnerable block or function.

\noindent\textit{\textbf{Obtaining explanation for suggested patch.}} After each patch-generation, we ask the LLM in a follow-up prompt to provide a brief natural-language explanation of the suggested patch. The resulting explanation $E$ is used for reasoning-based evaluation.

\subsection{Patch Evaluation for \system}\label{subsec:patch_eval}
Evaluating patches for logical vulnerabilities poses unique challenges. Although automated patch validation techniques exist \cite{tan2016anti,xin2017leveraging, le2023invalidator}, most rely on static code features or dynamic testing with predefined testcases or known exploits. Since logical vulnerabilities do not follow consistent patterns for identification or repair, extracting features that capture their essence is difficult, limiting the applicability of existing automated patch evaluation techniques. Learning-based approaches have also been proposed \cite{tian2022change,tian2023best}, but the lack of large, high-quality datasets of logical vulnerability patches constrains their effectiveness \cite{li2025sok}. More recently, LLM-based validation techniques \cite{zhou2024leveraging} have emerged, yet their reliability remains questionable due to difficulty in handling complex constraints \cite{li2025sok,liu2024exploring}. Several works employ CodeQL queries \cite{codeql2021} for testing \cite{sandoval2023lost, pearce2025asleep, pearce2023examining}, but crafting queries that capture logical vulnerabilities is non-trivial. Consequently, manual inspection remains the most dependable way to confirm that a proposed patch addresses the underlying logical vulnerability without introducing side effects \cite{li2025sok,pearce2025asleep,sandoval2023lost}. Based on these considerations, \system evaluates \textit{compilation and testing} and \textit{reasoning}.

\subsubsection{Compilation and Testing} \label{subsubsec:compilation_and_testing}
\system first checks whether a suggested patch is \emph{compilable} using the compilation script $C$. Because we focus on the core fix, we graft the suggested patch $Z$ onto the localized function $F_f$ or block $F_b$ over the ground-truth \emph{fixed} implementation $F$, ensuring that auxiliary modifications needed for compilation and testing are present while focusing only on the core fix. If compilation succeeds, the patch is \emph{compilable}. If a testing script $T$ is available, \system runs it on compilable patches; a patch that passes all tests is \emph{plausible}. However, a plausible patch may still be incorrect due to overfitting. Therefore, we manually inspect all plausible patches to determine whether they are \emph{correct}, i.e., whether they eliminate the vulnerability and introduce no unintended side effects. A patch that is not plausible cannot be correct.

\subsubsection{Reasoning}\label{subsubsec:reasoning}
Aside from compilation and testing, analyzing the steps a patch performs to fix the vulnerability is essential for logical vulnerabilities, especially to assess how \emph{reasonable} the patch is as a suggestion to the developer. To address the lack of an objective automated technique, we use LLM-assisted reasoning metrics similar to prior work on vulnerability identification \cite{ullah2023can}. Since LLMs understand program semantics well, they can identify similarity in rationale between a suggested patch and a ground-truth patch. A high similarity indicates that the candidate patch’s rationale aligns closely with the ground-truth, which we use as a proxy for \emph{reasonable} patches.

Concretely, \system compares the natural-language explanation $E$ of each candidate patch against a reference explanation $E_g$ of the ground-truth fix. We first prompt a ``judge'' LLM in a zero-shot setting with the vulnerability description $D$, vulnerable block $V_b$, and ground-truth fix block $F_b$ to derive $E_g$. When the patch-generating LLM outputs a suggested fix, it also provides $E$ (\S\ref{subsubsec:patch_generation_grafting}). We then compare $E$ and $E_g$ using cosine similarity ($CS$) \cite{ullah2023can}, and use a judge LLM \cite{ullah2023can} to provide a binary verdict ($J$) on whether they are similar. We also evaluate these reasoning metrics in Section \ref{sec:appendix:effectiveness_eval_metrics}.
\section{\dataset: Logical Vulnerability Dataset}\label{sec:dataset}

Previous AVR approaches curated different benchmark datasets for different programming languages such as Java \cite{just2014defects4j,bui2022vul4j,lin2017quixbugs}, C/C++ \cite{gao2021beyond,nikitopoulos2021crossvul,fan2020ac} and Python \cite{jimenez2023swe}.
However, we observe that these datasets contain very few logical flaws, and most have no clear security impact (e.g., the changing-offset issue in the Time-3 Defects4J sample \cite{time3defects4j}).

We, therefore, focus on security-relevant logical vulnerabilities and build \dataset, a new dataset containing 61 real-world cases. We identify vulnerabilities caused by deviations from expected behavior that undermine security guarantees by examining CVEs across 28 popular open-source implementations. For each sample, we use the CVE description as $D$, the fix commit as the fixed version $F$, and its parent commit as the vulnerable version $V$. We manually localize the core-fix vulnerable function and block ($V_f$, $V_b$), context $V_{ctx}$ (\S\ref{subsubsec:inputs_patch_generation}), and the corresponding fixed function and block ($F_f$, $F_b$). We create a compilation script $C$, extract relevant specification text $S$ when available, and collect testcases to build a test script $T$ where applicable. 
All these steps were performed manually to ensure the correctness and reliability of the ground truth dataset. Constructing each vulnerability data point took $\sim$10 person-hours.
To obtain a repair description $R$, we use LLMs by providing $D$, $V_b$, and $F_b$, asking for a natural-language description of the repair steps, and then manually refining it.

\noindent\textbf{Synthetic Java Samples.} Many state-of-the-art repair techniques~\cite{jiang2018shaping, jiang2023knod, campos2012gzoltar} are tailored to Java constructs and adopt Java-based fault localization, driven by the popularity of Defects4J~\cite{time3defects4j} and Vul4J~\cite{bui2022vul4j}. This limits their direct applicability to other languages (e.g., C/C++). To facilitate logical vulnerability repair research and benchmark Java-focused approaches \cite{jiang2018shaping, jiang2023knod, kulsum2024case}, we also construct 61 synthetic Java samples derived from our real-world cases, each including $V$, $F$, and testcases. Curation details are provided in Appendix \ref{sec:appendix:curation_of_dataset}.

\section{Evaluations \& Observations}\label{sec:experiment}

In this Section, we present the results and key findings from our \system evaluation. We analyze baseline AVR approaches and also off-the-shelf LLM prompting by varying \emph{source code}, \emph{auxiliary information}, and \emph{LLM configuration} on both real-world and synthetic (\dataset) vulnerabilities. Note that, for each claim presented in this Section, we also performed statistical significance tests for their justification, which are provided in Appendix \ref{sec:appendix:statistical_significance}. The code for \system, \dataset, and the generated patch suggestions are publicly available in \cite{logiceval}.

\subsection{Performance of Baseline Approaches}\label{subsec:baseline_performance}

\begin{table}[ht]
    \centering
    \tiny
    \setlength{\tabcolsep}{4pt}
    \begin{tabular}{cl|cccccc}
    \toprule
    \textbf{Dataset} & \textbf{Tool} & \textbf{C} & \textbf{T} & \textbf{CSL} & \textbf{CSQ} & \textbf{JL} & \textbf{JQ} \\
    \midrule
    \textbf{Real} & \textbf{VRPilot} & \cellcolor[RGB]{46,151,78}0.82 & \cellcolor[RGB]{218,241,212}0.08 & \cellcolor[RGB]{159,216,154}0.65 & \cellcolor[RGB]{144,209,141}0.66 & \cellcolor[RGB]{247,252,245}0.02 & \cellcolor[RGB]{238,249,234}0.07 \\
    \midrule
    \multirow{3}{*}{\textbf{Synthetic}} & \textbf{SimFix} & \cellcolor[RGB]{247,252,245}0.01 & \cellcolor[RGB]{247,252,245}0.00 & \cellcolor[RGB]{179,225,172}0.62 & \cellcolor[RGB]{159,216,154}0.64 & \cellcolor[RGB]{247,252,245}0.01 & \cellcolor[RGB]{247,252,245}0.00 \\
     & \textbf{KNOD} & \cellcolor[RGB]{182,226,176}0.35 & \cellcolor[RGB]{247,252,245}0.00 & \cellcolor[RGB]{165,219,159}0.64 & \cellcolor[RGB]{151,212,147}0.65 & \cellcolor[RGB]{247,252,245}0.02 & \cellcolor[RGB]{247,252,245}0.00 \\
     & \textbf{VRPilot} & \cellcolor[RGB]{101,189,110}0.56 & \cellcolor[RGB]{212,239,206}0.09 & \cellcolor[RGB]{159,216,154}0.65 & \cellcolor[RGB]{151,212,147}0.65 & \cellcolor[RGB]{233,247,229}0.15 & \cellcolor[RGB]{244,251,241}0.03 \\
    \bottomrule
    \end{tabular}
    
    \caption{Baseline tool results. ``C'' and ``T'' represent the \% of generated patches successfully compiled and passed all tests, respectively. ``CS'' represents average cosine similarity among suggested patches and ground-truth explanations, ``J'' represents the \% of patches a judging LLM determined similar to ground-truth fix. Suffixes ``L'', ``Q'' after each ``CS'' and ``J'' represent judging LLM, Llama, and Qwen, respectively. Color heat is relative to all results for the same metric across all our experiments.}
    \label{tab:baseline_results}
\end{table}

This experiment evaluates the performance, general capabilities, and limitations of existing automatic vulnerability repair (AVR) approaches. We use three state-of-the-art approaches to generate patches for logical vulnerabilities in \dataset: SimFix \cite{jiang2018shaping}, a non-learning-based method; KNOD \cite{jiang2023knod}, a deep learning-based technique; and VRPilot \cite{kulsum2024case}, an LLM-based program repair approach. Since SimFix and KNOD are Java-focused repair approaches, we only evaluate them on synthetic Java samples. On the other hand, we evaluate VRPilot on both real-world and synthetic samples of \dataset.  The detailed adoption process for the baselines is discussed in Appendix \ref{sec:appendix:adopt_baseline}.

We present the compilation and testing results, and automated reasoning scores of the baseline approaches in both real-world and synthetic examples in Table \ref{tab:baseline_results}. VRPilot, being an LLM-based approach, performs better compared to other baseline approaches. However, its CoT-based approach has a lower reasoning score than zero-shot prompting techniques, as we discuss in Section \ref{subsec:llm_performance}. Both SimFix and KNOD mostly generate unreasonable patches and use incorrect repair logic. E.g., patches from SimFix~\cite{jiang2018shaping} and from KNOD~\cite{jiang2023knod} generate totally unreasonable logic.

\subsection{Performance of Off-the-shelf LLMs}\label{subsec:llm_performance}

This experiment aims to evaluate the performance of off-the-shelf LLMs to repair logical vulnerabilities using \system. Specifically, we evaluate the performance of LLMs along three dimensions when prompting for logical vulnerabilities: \emph{LLM configuration}, \emph{source code}, and \emph{auxiliary information}. Examples for manual inspection are provided in  Appendix \ref{sec:appendix:example_codes}. 

\noindent\textbf{\textit{LLMs considered and default configurations.}} We leverage three popular off-the-shelf LLMs: Meta Llama 3.1~\cite{llama3}, Qwen 2.5~\cite{qwen}, and OpenAI o3-mini~\cite{o3mini}. 
For each experiment, each LLM is used to generate and evaluate patches using \system for each vulnerability. Note that our work aims not to benchmark all LLMs to repair logical vulnerabilities, but rather to identify the capabilities and limitations of LLMs in general towards repairing logical vulnerabilities. 

In each experiment, we adjust one dimension and set the others to their defaults. We use a temperature of 0.2, role orientation, and zero-shot prompts as the default LLM configuration, as they are the best-performing parameters (or at least equally performing compared to other choices) in their respective dimension as per our experiments described in detail below. Also, we use the vulnerable block $V_b$ as the default provided vulnerable source code and the vulnerability information $D$ as the default provided auxiliary information while varying configurations of other dimensions. We provide the prompt templates used in our experiments in  Appendix \ref{sec:appendix:prompts_used_llm}. For brevity, in this section, we display only the results for real-world samples in \dataset across our experiments. The corresponding results for synthetic examples in \dataset are shown in Appendix \ref{sec:appendix:synthetic_example_results}.

For each LLM, we leverage the other two LLMs as judge LLMs, and obtain the cosine similarity scores among explanations, and also judging verdicts (Section \ref{subsec:patch_eval}). We display the average cosine similarity scores and percentage of patches whose explanation aligned with the ground-truth fix according to each judge-LLM.

\begin{table}[ht]
    \centering
    \setlength{\tabcolsep}{0.8pt}
    \tiny
    \resizebox{\linewidth}{!}{
    \begin{tabular}{c|cccccc|cccccc|cccccc}
        \toprule
        \multicolumn{1}{c|}{} & \multicolumn{6}{c|}{\textbf{Llama-3.1}} & \multicolumn{6}{c|}{\textbf{Qwen-2.5}} & \multicolumn{6}{c}{\textbf{OpenAI-o3-mini}} \\
        \cmidrule(lr){2-7}\cmidrule(lr){8-13}\cmidrule(lr){14-19}
        \textbf{P} & \textbf{C} & \textbf{T} & \textbf{CSQ} & \textbf{CSO} & \textbf{JQ} & \textbf{JO} & \textbf{C} & \textbf{T} & \textbf{CSL} & \textbf{CSO} & \textbf{JL} & \textbf{JO} & \textbf{C} & \textbf{T} & \textbf{CSL} & \textbf{CSQ} & \textbf{JL} & \textbf{JQ} \\
        \midrule
        \textbf{P1} & \cellcolor[RGB]{118,197,120}0.50 & \cellcolor[RGB]{229,245,224}0.06 & \cellcolor[RGB]{53,158,83}0.81 & \cellcolor[RGB]{91,184,106}0.76 & \cellcolor[RGB]{235,248,231}0.10 & \cellcolor[RGB]{201,234,194}0.35 & \cellcolor[RGB]{69,173,95}0.66 & \cellcolor[RGB]{240,250,237}0.04 & \cellcolor[RGB]{54,159,84}0.81 & \cellcolor[RGB]{112,194,116}0.73 & \cellcolor[RGB]{142,209,140}0.60 & \cellcolor[RGB]{165,219,159}0.53 & \cellcolor[RGB]{93,185,107}0.58 & \cellcolor[RGB]{223,243,217}0.07 & \cellcolor[RGB]{62,169,91}0.77 & \cellcolor[RGB]{57,162,86}0.77 & \cellcolor[RGB]{83,180,102}0.79 & \cellcolor[RGB]{182,226,176}0.32 \\
        \textbf{P2} & \cellcolor[RGB]{127,201,127}0.48 & \cellcolor[RGB]{224,243,218}0.07 & \cellcolor[RGB]{53,158,83}0.81 & \cellcolor[RGB]{91,184,106}0.76 & \cellcolor[RGB]{231,246,226}0.13 & \cellcolor[RGB]{205,236,198}0.33 & \cellcolor[RGB]{58,163,87}0.70 & \cellcolor[RGB]{240,250,237}0.04 & \cellcolor[RGB]{54,159,84}0.81 & \cellcolor[RGB]{99,188,110}0.74 & \cellcolor[RGB]{158,216,152}0.54 & \cellcolor[RGB]{163,218,157}0.54 & -- & -- & -- & -- & -- & -- \\
        \midrule
        \textbf{P3} & \cellcolor[RGB]{99,188,110}0.54 & \cellcolor[RGB]{229,245,224}0.06 & \cellcolor[RGB]{53,158,83}0.81 & \cellcolor[RGB]{81,179,101}0.77 & \cellcolor[RGB]{231,246,226}0.13 & \cellcolor[RGB]{199,233,192}0.36 & \cellcolor[RGB]{77,177,99}0.64 & \cellcolor[RGB]{240,250,237}0.04 & \cellcolor[RGB]{61,168,90}0.80 & \cellcolor[RGB]{99,188,110}0.74 & \cellcolor[RGB]{158,216,152}0.54 & \cellcolor[RGB]{168,220,162}0.52 & -- & -- & -- & -- & -- & -- \\
        \textbf{P4} & \cellcolor[RGB]{118,197,120}0.50 & \cellcolor[RGB]{229,245,224}0.06 & \cellcolor[RGB]{53,158,83}0.81 & \cellcolor[RGB]{101,189,110}0.75 & \cellcolor[RGB]{234,247,230}0.11 & \cellcolor[RGB]{170,221,164}0.49 & \cellcolor[RGB]{69,173,95}0.66 & \cellcolor[RGB]{240,250,237}0.04 & \cellcolor[RGB]{61,168,90}0.80 & \cellcolor[RGB]{148,211,144}0.70 & \cellcolor[RGB]{170,221,164}0.49 & \cellcolor[RGB]{170,221,164}0.51 & \cellcolor[RGB]{59,166,89}0.66 & \cellcolor[RGB]{223,243,217}0.07 & \cellcolor[RGB]{52,157,83}0.79 & \cellcolor[RGB]{46,151,78}0.79 & \cellcolor[RGB]{128,202,128}0.64 & \cellcolor[RGB]{155,214,150}0.41 \\
        \midrule
        \textbf{P5} & \cellcolor[RGB]{104,190,112}0.53 & \cellcolor[RGB]{238,249,235}0.03 & \cellcolor[RGB]{46,151,78}0.82 & \cellcolor[RGB]{81,179,101}0.77 & \cellcolor[RGB]{231,246,226}0.13 & \cellcolor[RGB]{181,226,174}0.44 & \cellcolor[RGB]{102,190,111}0.57 & \cellcolor[RGB]{242,250,239}0.03 & \cellcolor[RGB]{61,168,90}0.80 & \cellcolor[RGB]{99,188,110}0.74 & \cellcolor[RGB]{165,219,159}0.51 & \cellcolor[RGB]{181,226,174}0.46 & \cellcolor[RGB]{55,160,85}0.68 & \cellcolor[RGB]{218,241,212}0.08 & \cellcolor[RGB]{62,169,91}0.77 & \cellcolor[RGB]{57,162,86}0.77 & \cellcolor[RGB]{122,199,123}0.66 & \cellcolor[RGB]{194,231,187}0.28 \\
        \textbf{P6} & \cellcolor[RGB]{99,188,110}0.54 & \cellcolor[RGB]{247,252,245}0.00 & \cellcolor[RGB]{46,151,78}0.82 & \cellcolor[RGB]{110,194,115}0.74 & \cellcolor[RGB]{226,244,220}0.16 & \cellcolor[RGB]{199,233,192}0.36 & \cellcolor[RGB]{77,177,99}0.64 & \cellcolor[RGB]{240,250,237}0.04 & \cellcolor[RGB]{61,168,90}0.80 & \cellcolor[RGB]{112,194,116}0.73 & \cellcolor[RGB]{165,219,159}0.51 & \cellcolor[RGB]{163,218,157}0.54 & \cellcolor[RGB]{83,180,102}0.60 & \cellcolor[RGB]{238,249,235}0.03 & \cellcolor[RGB]{247,252,245}0.47 & \cellcolor[RGB]{247,252,245}0.46 & \cellcolor[RGB]{209,237,203}0.30 & \cellcolor[RGB]{232,247,228}0.11 \\
        \textbf{P7} & \cellcolor[RGB]{139,207,137}0.45 & \cellcolor[RGB]{247,252,245}0.00 & \cellcolor[RGB]{131,203,130}0.73 & \cellcolor[RGB]{101,189,110}0.75 & \cellcolor[RGB]{234,247,230}0.11 & \cellcolor[RGB]{203,235,196}0.34 & \cellcolor[RGB]{91,184,106}0.60 & \cellcolor[RGB]{240,250,237}0.04 & \cellcolor[RGB]{121,199,122}0.75 & \cellcolor[RGB]{62,169,91}0.77 & \cellcolor[RGB]{145,210,142}0.59 & \cellcolor[RGB]{176,224,170}0.48 & \cellcolor[RGB]{122,199,123}0.51 & \cellcolor[RGB]{212,239,206}0.09 & \cellcolor[RGB]{86,182,103}0.74 & \cellcolor[RGB]{94,186,107}0.72 & \cellcolor[RGB]{106,191,113}0.72 & \cellcolor[RGB]{164,219,158}0.38 \\
        \textbf{P8} & \cellcolor[RGB]{46,151,78}0.68 & \cellcolor[RGB]{236,248,232}0.04 & \cellcolor[RGB]{177,224,171}0.68 & \cellcolor[RGB]{162,218,156}0.68 & \cellcolor[RGB]{244,251,241}0.03 & \cellcolor[RGB]{239,249,236}0.10 & \cellcolor[RGB]{46,151,78}0.75 & \cellcolor[RGB]{238,249,235}0.05 & \cellcolor[RGB]{224,243,218}0.64 & \cellcolor[RGB]{214,239,208}0.63 & \cellcolor[RGB]{242,251,240}0.07 & \cellcolor[RGB]{244,251,241}0.07 & \cellcolor[RGB]{59,166,89}0.66 & \cellcolor[RGB]{218,241,212}0.08 & \cellcolor[RGB]{173,222,166}0.63 & \cellcolor[RGB]{173,222,166}0.62 & \cellcolor[RGB]{218,241,212}0.25 & \cellcolor[RGB]{234,247,230}0.10 \\
        \bottomrule
        \end{tabular}
    }
    \caption{Compilation, testing, and reasoning score results for different LLM configurations in real-world vulnerabilities. Column meanings are the same as in Table \ref{tab:baseline_results}. ``P'' represents the prompt IDs. Suffixes ``L'', ``Q'', and ``O'' after each ``CS'' and ``J'' represent judging LLM Llama, Qwen, and OpenAI-o3-mini, respectively.}
    \label{tab:llm_config_real}
\end{table}
\subsubsection{Impact of Adjusting LLM configurations}\label{subsubsec:impact_of_adjusting_llm_conf}
To assess the impact of LLM configurations on patch generation performance, we adjust the temperature, orientation, and prompting strategy of LLMs while keeping the prompt contents the same. We provide the vulnerable block $V_b$, the vulnerability description $D$, and ask the LLM to suggest a patch using \system. To assess temperature, we run patch generation with temperatures 0.2 (low), 0.5 (medium), and 0.9 (high) using prompts \textbf{P1}, \textbf{P2}, and \textbf{P3}, respectively. To assess orientation, we adopt a task-oriented prompt in \textbf{P4}. Finally to assess prompting strategy, we consider: (i) zero-shot prompt \textbf{P5} with $V_b$ and $D$; (ii) few-shot prompt \textbf{P6}, where in addition to $V_b$ and $D$, we provide a real-world sample from \system with its $D$, $V_b$, and ground-truth $F_b$ as an example fix; (iii) chain-of-thought (CoT) prompt \textbf{P7}, where we first provide $V_b$ and $D$ and ask for steps $\mathsf{R_{suggested}}$ to repair the vulnerability, then provide $V_b$ and $\mathsf{R_{suggested}}$ to obtain a patch; and (iv) CoT prompt \textbf{P8}, similar to \textbf{P7} except we do not include $\mathsf{R_{suggested}}$ in the second prompt.

We present the compilation and testing results, along with the average reasoning scores, in Table \ref{tab:llm_config_real} for real-world examples in \dataset. We observe that the compilation, testing, and reasoning scores for \textbf{P1}, \textbf{P2}, and \textbf{P3} are similar (we cannot adjust the temperature of openAI-o3-mini), indicating \textit{\textbf{there is minimal effect of logical vulnerability repair performance based on adjusting temperature.}} Similarly, the scores for prompt \textbf{P4} are also similar to \textbf{P1}, indicating \textit{\textbf{there is minimal effect of logical vulnerability repair performance based on orientation as well.}}


\newtcolorbox{takeawaybox}[1][]{
  enhanced,
  colback=gray!10,        
  colframe=blue!50!black,   
  boxrule=0.5pt,
  arc=1mm,
  left=3pt,
  right=3pt,
  top=3pt,
  bottom=3pt,
  fontupper=\small,
  fonttitle=\bfseries,
  title={#1}
}


When evaluating different prompting strategies, we observe that \textit{\textbf{usually zero-shot prompt \textbf{P5} obtains higher compilation percentage and reasoning scores than CoT prompts, especially on real-world examples.}} Several compilation errors occur for CoT prompt \textbf{P7} because it creates extra undefined variables based on the reasoning suggestion from its first prompt (Listing ~\ref{tab:listing_1}). Prompt \textbf{P8} omits all auxiliary information and relies only on the vulnerable source code ($V_b$). As a result, it generates more compilable patches but achieves significantly lower reasoning scores than both \textbf{P5} and \textbf{P7}. This indicates \textit{\textbf{auxiliary information is required alongside the source code to achieve reasonable patches}}. Scores for few-shot prompt \textbf{P6} are usually marginally lower than zero-shot prompts across the three LLMs; we manually did not observe any contrasting behavior pattern between the two strategies.

\begin{table}[ht]
    \centering
    \setlength{\tabcolsep}{0.8pt}
    \tiny
    \resizebox{\linewidth}{!}{
    \begin{tabular}{c|cccccc|cccccc|cccccc}
        \toprule
        \multicolumn{1}{c|}{} & \multicolumn{6}{c|}{\textbf{Llama-3.1}} & \multicolumn{6}{c|}{\textbf{Qwen-2.5}} & \multicolumn{6}{c}{\textbf{OpenAI-o3-mini}} \\
        \cmidrule(lr){2-7}\cmidrule(lr){8-13}\cmidrule(lr){14-19}
        \textbf{P} & \textbf{C} & \textbf{T} & \textbf{CSQ} & \textbf{CSO} & \textbf{JQ} & \textbf{JO} & \textbf{C} & \textbf{T} & \textbf{CSL} & \textbf{CSO} & \textbf{JL} & \textbf{JO} & \textbf{C} & \textbf{T} & \textbf{CSL} & \textbf{CSQ} & \textbf{JL} & \textbf{JQ} \\
        \midrule
        \textbf{P9} & \cellcolor[RGB]{104,190,112}0.53 & \cellcolor[RGB]{224,243,218}0.07 & \cellcolor[RGB]{46,151,78}0.82 & \cellcolor[RGB]{91,184,106}0.76 & \cellcolor[RGB]{237,249,234}0.08 & \cellcolor[RGB]{188,229,181}0.41 & \cellcolor[RGB]{91,184,106}0.60 & \cellcolor[RGB]{234,247,230}0.07 & \cellcolor[RGB]{54,159,84}0.81 & \cellcolor[RGB]{99,188,110}0.74 & \cellcolor[RGB]{158,216,152}0.54 & \cellcolor[RGB]{175,223,169}0.49 & \cellcolor[RGB]{65,171,93}0.64 & \cellcolor[RGB]{182,226,176}0.14 & \cellcolor[RGB]{57,162,86}0.78 & \cellcolor[RGB]{52,157,83}0.78 & \cellcolor[RGB]{110,194,115}0.70 & \cellcolor[RGB]{185,227,178}0.31 \\
        \textbf{P10} & \cellcolor[RGB]{94,186,107}0.55 & \cellcolor[RGB]{224,243,218}0.07 & \cellcolor[RGB]{78,178,99}0.78 & \cellcolor[RGB]{91,184,106}0.76 & \cellcolor[RGB]{245,252,243}0.02 & \cellcolor[RGB]{203,235,196}0.34 & \cellcolor[RGB]{72,175,96}0.65 & \cellcolor[RGB]{232,247,228}0.08 & \cellcolor[RGB]{61,168,90}0.80 & \cellcolor[RGB]{99,188,110}0.74 & \cellcolor[RGB]{158,216,152}0.54 & \cellcolor[RGB]{154,214,149}0.58 & \cellcolor[RGB]{46,151,78}0.71 & \cellcolor[RGB]{169,221,163}0.16 & \cellcolor[RGB]{77,177,99}0.75 & \cellcolor[RGB]{69,173,95}0.75 & \cellcolor[RGB]{137,206,135}0.61 & \cellcolor[RGB]{170,221,164}0.36 \\
        \textbf{P11} & \cellcolor[RGB]{122,199,123}0.49 & \cellcolor[RGB]{229,245,224}0.06 & \cellcolor[RGB]{53,158,83}0.81 & \cellcolor[RGB]{91,184,106}0.76 & \cellcolor[RGB]{235,248,231}0.10 & \cellcolor[RGB]{183,227,177}0.43 & \cellcolor[RGB]{83,180,102}0.62 & \cellcolor[RGB]{225,244,219}0.11 & \cellcolor[RGB]{46,151,78}0.82 & \cellcolor[RGB]{86,182,103}0.75 & \cellcolor[RGB]{170,221,164}0.49 & \cellcolor[RGB]{175,223,169}0.49 & \cellcolor[RGB]{55,160,85}0.68 & \cellcolor[RGB]{182,226,176}0.14 & \cellcolor[RGB]{62,169,91}0.77 & \cellcolor[RGB]{57,162,86}0.77 & \cellcolor[RGB]{106,191,113}0.72 & \cellcolor[RGB]{170,221,164}0.36 \\
        \bottomrule
        \end{tabular}
    }
    \caption{Compilation, testing, and reasoning score results for varying portions of the source code provided in real-world vulnerabilities. Column meanings are the same as in Table \ref{tab:baseline_results} and \ref{tab:llm_config_real}.}
    \label{tab:source_code_real}
\end{table}
\vspace{-2mm}

\subsubsection{Impacts of Adjusting Source Code}\label{subsubsec:impact_of_adjusting_source_code} 
We evaluate the advantages and disadvantages of providing different portions of vulnerable source code that are manually localized (See Section \ref{subsubsec:vulnerability_localization}). 
We either provide $V_b$, where the core fix is required, or $V_f$. Providing $V_b$ allows more fine-grained localization, whereas $V_f$ provides broader context. 
We first compare two prompting strategies: \textbf{P9}, which supplies only the vulnerable block $V_b$, and \textbf{P10}, which supplies the entire vulnerable function $V_f$. Across both real‐world and synthetic datasets—and for all LLMs tested, \textbf{P10} achieves marginally higher compilation and testing success rates, while \textbf{P9} has marginally better reasoning scores (the $J$ score differences are not statistically significant, however, the $CS$ scores are, see Appendix \ref{sec:appendix:statistical_significance}).

Manual inspection reveals that when only the vulnerable block is provided, LLMs sometimes fail to resolve the correct variable names and instead introduce placeholder variables, leading to compilation failures. Supplying the full function context may mitigate the need for placeholder variables. Conversely, when functions are very large, giving the entire function can make it harder for the LLMs to locate the exact block that needs to be modified, occasionally resulting in unreasonable patch suggestions (as in Listing ~\ref{tab:listing_3}), suggesting that \textbf{\textit{providing only a vulnerable block without context sometimes leads LLMs to create placeholder variables. In contrast, when the entire vulnerable function is provided, LLMs sometimes fail to locate the block to modify to fix the vulnerability.}}


We also consider another prompting strategy \textbf{P11}, which provides \emph{context} (Section \ref{subsubsec:inputs_patch_generation}) and compare the performance with \textbf{P9}. We observe a marginal increase in compilation pass rate across all LLMs and sample types (both real-world and synthetic) except for Llama patches over real-world samples. In contrast, reasoning scores do not improve much. This suggests that \textit{\textbf{providing context somewhat facilitates the compilation process.}}

\begin{table}[ht]
    \centering
    \setlength{\tabcolsep}{0.8pt}
    \tiny
    \resizebox{\linewidth}{!}{
    \begin{tabular}{c|cccccc|cccccc|cccccc}
        \toprule
        \multicolumn{1}{c|}{} & \multicolumn{6}{c|}{\textbf{Llama-3.1}} & \multicolumn{6}{c|}{\textbf{Qwen-2.5}} & \multicolumn{6}{c}{\textbf{OpenAI-o3-mini}} \\
        \cmidrule(lr){2-7}\cmidrule(lr){8-13}\cmidrule(lr){14-19}
        \textbf{P} & \textbf{C} & \textbf{T} & \textbf{CSQ} & \textbf{CSO} & \textbf{JQ} & \textbf{JO} & \textbf{C} & \textbf{T} & \textbf{CSL} & \textbf{CSO} & \textbf{JL} & \textbf{JO} & \textbf{C} & \textbf{T} & \textbf{CSL} & \textbf{CSQ} & \textbf{JL} & \textbf{JQ} \\
        \midrule
        \textbf{P12} & \cellcolor[RGB]{52,157,83}0.66 & \cellcolor[RGB]{236,248,232}0.04 & \cellcolor[RGB]{186,228,179}0.67 & \cellcolor[RGB]{198,233,191}0.63 & \cellcolor[RGB]{245,252,243}0.02 & \cellcolor[RGB]{239,249,236}0.10 & \cellcolor[RGB]{46,151,78}0.75 & \cellcolor[RGB]{238,249,235}0.05 & \cellcolor[RGB]{234,247,230}0.62 & \cellcolor[RGB]{229,245,224}0.61 & \cellcolor[RGB]{238,249,235}0.11 & \cellcolor[RGB]{237,249,234}0.13 & \cellcolor[RGB]{83,180,102}0.60 & \cellcolor[RGB]{223,243,217}0.07 & \cellcolor[RGB]{165,219,159}0.64 & \cellcolor[RGB]{159,216,154}0.64 & \cellcolor[RGB]{227,245,222}0.20 & \cellcolor[RGB]{238,249,234}0.07 \\
        \textbf{P13} & \cellcolor[RGB]{94,186,107}0.55 & \cellcolor[RGB]{238,249,235}0.03 & \cellcolor[RGB]{53,158,83}0.81 & \cellcolor[RGB]{91,184,106}0.76 & \cellcolor[RGB]{234,247,230}0.11 & \cellcolor[RGB]{188,229,181}0.41 & \cellcolor[RGB]{91,184,106}0.60 & \cellcolor[RGB]{240,250,237}0.04 & \cellcolor[RGB]{61,168,90}0.80 & \cellcolor[RGB]{112,194,116}0.73 & \cellcolor[RGB]{145,210,142}0.59 & \cellcolor[RGB]{176,224,170}0.48 & \cellcolor[RGB]{65,171,93}0.64 & \cellcolor[RGB]{182,226,176}0.14 & \cellcolor[RGB]{62,169,91}0.77 & \cellcolor[RGB]{57,162,86}0.77 & \cellcolor[RGB]{110,194,115}0.70 & \cellcolor[RGB]{164,219,158}0.38 \\
        \textbf{P14} & \cellcolor[RGB]{122,199,123}0.49 & \cellcolor[RGB]{247,252,245}0.00 & \cellcolor[RGB]{53,158,83}0.81 & \cellcolor[RGB]{72,175,96}0.78 & \cellcolor[RGB]{222,242,216}0.18 & \cellcolor[RGB]{165,219,159}0.51 & \cellcolor[RGB]{110,194,115}0.55 & \cellcolor[RGB]{247,252,245}0.00 & \cellcolor[RGB]{61,168,90}0.80 & \cellcolor[RGB]{99,188,110}0.74 & \cellcolor[RGB]{118,197,120}0.69 & \cellcolor[RGB]{138,207,136}0.64 & \cellcolor[RGB]{122,199,123}0.51 & \cellcolor[RGB]{212,239,206}0.09 & \cellcolor[RGB]{57,162,86}0.78 & \cellcolor[RGB]{57,162,86}0.77 & \cellcolor[RGB]{106,191,113}0.72 & \cellcolor[RGB]{162,218,156}0.39 \\
        \textbf{P15} & \cellcolor[RGB]{63,170,92}0.62 & \cellcolor[RGB]{224,243,218}0.07 & \cellcolor[RGB]{53,158,83}0.81 & \cellcolor[RGB]{81,179,101}0.77 & \cellcolor[RGB]{149,212,145}0.46 & \cellcolor[RGB]{109,193,114}0.72 & \cellcolor[RGB]{58,163,87}0.70 & \cellcolor[RGB]{232,247,228}0.08 & \cellcolor[RGB]{54,159,84}0.81 & \cellcolor[RGB]{86,182,103}0.75 & \cellcolor[RGB]{88,182,104}0.79 & \cellcolor[RGB]{102,190,111}0.77 & \cellcolor[RGB]{106,191,113}0.55 & \cellcolor[RGB]{189,229,183}0.13 & \cellcolor[RGB]{69,173,95}0.76 & \cellcolor[RGB]{62,169,91}0.76 & \cellcolor[RGB]{67,172,94}0.84 & \cellcolor[RGB]{97,187,109}0.57 \\
        \textbf{P16} & \cellcolor[RGB]{60,167,89}0.63 & \cellcolor[RGB]{247,252,245}0.00 & \cellcolor[RGB]{89,183,105}0.77 & \cellcolor[RGB]{110,194,115}0.74 & \cellcolor[RGB]{187,228,180}0.33 & \cellcolor[RGB]{115,196,118}0.70 & \cellcolor[RGB]{80,179,100}0.63 & \cellcolor[RGB]{247,252,245}0.00 & \cellcolor[RGB]{132,204,131}0.74 & \cellcolor[RGB]{124,200,124}0.72 & \cellcolor[RGB]{124,200,124}0.67 & \cellcolor[RGB]{129,203,129}0.67 & \cellcolor[RGB]{122,199,123}0.51 & \cellcolor[RGB]{218,241,212}0.08 & \cellcolor[RGB]{104,190,112}0.72 & \cellcolor[RGB]{104,190,112}0.71 & \cellcolor[RGB]{106,191,113}0.72 & \cellcolor[RGB]{144,209,141}0.44 \\
        \bottomrule
        \end{tabular}
    }
    \caption{Compilation, testing, and reasoning scores for varying auxiliary information provided in prompts for real-world vulnerabilities. Column meanings are the same as in Table \ref{tab:baseline_results} and \ref{tab:llm_config_real}.}
    \label{tab:aux_info_real}
\end{table}

\subsubsection{Impacts of Adjusting Auxiliary Information} \label{subsubsec:impact_adjusting_aux_information}
The goal of this experiment is to observe to what extent different auxiliary information facilitates repairing logical vulnerabilities with LLMs. Hence, we consider different combinations of auxiliary information in each of the five different prompts: prompt \textbf{P12} having no auxiliary information (i.e., it only provides the vulnerable block $V_b$ to LLMs and ask for a patch), prompt \textbf{P13} provides only the vulnerability information $D$, prompt \textbf{P14} additionally provides a specification text (if available, if not this defaults to \textbf{P13}), prompt \textbf{P15} provides the repair steps $R$ alongside the vulnerability information $D$, and prompt \textbf{P16} only provides the repair steps $R$, without the vulnerability information.

The compilation, testing, and reasoning results are shown in Table \ref{tab:aux_info_real} for real-world samples. Here, first, we observe that prompt \textbf{P12} has a much higher compilation rate, but much lower reasoning scores compared to other prompts. Upon manual inspection, we find that without any auxiliary information, LLM tends to treat the vulnerable source code as a possible memory vulnerability issue (as in this use-after-free example provided in Listing \ref{tab:listing_5}) and suggests patches to fix those vulnerabilities. These patches pass compilation, but do not meaningfully attempt to repair the logical vulnerability. This suggests that \textit{\textbf{Auxiliary information about the vulnerability (e.g., vulnerability description, specification text, or repair steps) is essential for generating accurate patches for logical vulnerabilities.}}

Furthermore, we observed that in some cases, since specification texts usually describe expected behavior in a different abstraction from the implementation, LLMs tend to create undeclared variables adopted from the text (example provided in Table \ref{tab:listing_6}), which is a similar phenomenon we observed while evaluating prompt \textbf{P7}. Consequently, Prompt \textbf{P13} has much higher reasoning scores compared to prompt \textbf{P12}, and prompt \textbf{P14} has similar reasoning scores but marginally less compilation rate compared to prompt \textbf{P13}.


\subsubsection{Overall Observations}\label{subsubsec:observations}
Off-the-shelf LLMs perform significantly better when provided auxiliary information, outperforming other baseline approaches. However, the overall performance leaves room for improvement, especially in a real-world environment where the LLMs generated a correct patch for only 5 out of 61 samples. However, the observations suggest that repair prompts should incorporate localization cues and more relevant context about the vulnerability. To reduce hallucinations, any text included in the prompt should be highly specific and aligned with the actual implementation.

\begin{table}[ht]
    \centering
    \setlength{\tabcolsep}{0.8pt}
    \tiny
    \resizebox{\linewidth}{!}{
    \begin{tabular}{c|cccccc|cccccc|cccccc}
        \toprule
        \multicolumn{1}{c|}{} & \multicolumn{6}{c|}{\textbf{Llama-3.1}} & \multicolumn{6}{c|}{\textbf{Qwen-2.5}} & \multicolumn{6}{c}{\textbf{OpenAI-o3-mini}} \\
        \cmidrule(lr){2-7}\cmidrule(lr){8-13}\cmidrule(lr){14-19}
        \textbf{P} & \textbf{C} & \textbf{T} & \textbf{CSQ} & \textbf{CSO} & \textbf{JQ} & \textbf{JO} & \textbf{C} & \textbf{T} & \textbf{CSL} & \textbf{CSO} & \textbf{JL} & \textbf{JO} & \textbf{C} & \textbf{T} & \textbf{CSL} & \textbf{CSQ} & \textbf{JL} & \textbf{JQ} \\
        \midrule
        \textbf{P17} & \cellcolor[RGB]{114,195,117}0.51 & \cellcolor[RGB]{238,249,235}0.03 & \cellcolor[RGB]{247,252,245}0.56 & \cellcolor[RGB]{247,252,245}0.52 & \cellcolor[RGB]{245,252,243}0.02 & \cellcolor[RGB]{242,251,240}0.07 & \cellcolor[RGB]{168,220,162}0.38 & \cellcolor[RGB]{247,252,245}0.00 & \cellcolor[RGB]{244,251,241}0.60 & \cellcolor[RGB]{243,251,240}0.58 & \cellcolor[RGB]{247,252,245}0.02 & \cellcolor[RGB]{242,251,240}0.08 & \cellcolor[RGB]{199,233,192}0.30 & \cellcolor[RGB]{238,249,235}0.03 & \cellcolor[RGB]{214,239,208}0.56 & \cellcolor[RGB]{214,239,208}0.55 & \cellcolor[RGB]{247,252,245}0.02 & \cellcolor[RGB]{247,252,245}0.00 \\
        \textbf{P18} & \cellcolor[RGB]{148,211,144}0.43 & \cellcolor[RGB]{247,252,245}0.00 & \cellcolor[RGB]{214,239,208}0.63 & \cellcolor[RGB]{221,242,215}0.59 & \cellcolor[RGB]{247,252,245}0.00 & \cellcolor[RGB]{242,251,240}0.07 & \cellcolor[RGB]{168,220,162}0.38 & \cellcolor[RGB]{247,252,245}0.00 & \cellcolor[RGB]{244,251,241}0.60 & \cellcolor[RGB]{247,252,245}0.57 & \cellcolor[RGB]{245,251,242}0.05 & \cellcolor[RGB]{244,251,241}0.07 & \cellcolor[RGB]{218,241,212}0.23 & \cellcolor[RGB]{232,246,227}0.05 & \cellcolor[RGB]{185,227,178}0.61 & \cellcolor[RGB]{179,225,172}0.61 & \cellcolor[RGB]{245,251,242}0.05 & \cellcolor[RGB]{245,252,243}0.02 \\
        \textbf{P19} & \cellcolor[RGB]{148,211,144}0.43 & \cellcolor[RGB]{247,252,245}0.00 & \cellcolor[RGB]{221,242,215}0.62 & \cellcolor[RGB]{221,242,215}0.59 & \cellcolor[RGB]{244,251,241}0.03 & \cellcolor[RGB]{241,250,238}0.08 & \cellcolor[RGB]{168,220,162}0.38 & \cellcolor[RGB]{247,252,245}0.00 & \cellcolor[RGB]{239,249,236}0.61 & \cellcolor[RGB]{234,247,230}0.60 & \cellcolor[RGB]{245,251,242}0.05 & \cellcolor[RGB]{241,250,238}0.10 & \cellcolor[RGB]{232,246,227}0.17 & \cellcolor[RGB]{247,252,245}0.00 & \cellcolor[RGB]{185,227,178}0.61 & \cellcolor[RGB]{185,227,178}0.60 & \cellcolor[RGB]{239,249,236}0.10 & \cellcolor[RGB]{245,252,243}0.02 \\
        \textbf{P20} & \cellcolor[RGB]{152,213,148}0.42 & \cellcolor[RGB]{247,252,245}0.00 & \cellcolor[RGB]{161,217,155}0.70 & \cellcolor[RGB]{176,224,170}0.66 & \cellcolor[RGB]{239,249,236}0.07 & \cellcolor[RGB]{233,247,228}0.16 & \cellcolor[RGB]{168,220,162}0.38 & \cellcolor[RGB]{247,252,245}0.00 & \cellcolor[RGB]{224,243,218}0.64 & \cellcolor[RGB]{214,239,208}0.63 & \cellcolor[RGB]{213,239,207}0.28 & \cellcolor[RGB]{230,246,225}0.20 & \cellcolor[RGB]{199,233,192}0.30 & \cellcolor[RGB]{247,252,245}0.00 & \cellcolor[RGB]{128,202,128}0.69 & \cellcolor[RGB]{112,194,116}0.70 & \cellcolor[RGB]{202,234,195}0.34 & \cellcolor[RGB]{241,250,238}0.05 \\
        \textbf{P21} & \cellcolor[RGB]{174,223,167}0.36 & \cellcolor[RGB]{247,252,245}0.00 & \cellcolor[RGB]{59,166,89}0.80 & \cellcolor[RGB]{101,189,110}0.75 & \cellcolor[RGB]{235,248,231}0.10 & \cellcolor[RGB]{173,222,166}0.48 & \cellcolor[RGB]{168,220,162}0.38 & \cellcolor[RGB]{247,252,245}0.00 & \cellcolor[RGB]{132,204,131}0.74 & \cellcolor[RGB]{159,216,154}0.69 & \cellcolor[RGB]{124,200,124}0.67 & \cellcolor[RGB]{138,207,136}0.64 & \cellcolor[RGB]{223,243,217}0.21 & \cellcolor[RGB]{247,252,245}0.00 & \cellcolor[RGB]{77,177,99}0.75 & \cellcolor[RGB]{69,173,95}0.75 & \cellcolor[RGB]{96,186,108}0.75 & \cellcolor[RGB]{207,236,200}0.23 \\
        \bottomrule
        \end{tabular}
    }
    \caption{Compilation, testing, and reasoning score results using templates provided in \cite{pearce2023examining} in real-world vulnerabilities. Column meanings are the same as in Table \ref{tab:baseline_results} and \ref{tab:llm_config_real}.}
    \label{tab:existing_technique_real}
\end{table}

\subsection{Comparison with Other Evaluation Works}\label{subsec:comparison}

Among existing evaluation frameworks for vulnerability repair, Pearce et al. \cite{pearce2023examining} propose a framework for evaluating LLM-generated patches for security vulnerability repair. We use their prompt templates (Table IV in \cite{pearce2023examining}) to generate patches for logical vulnerabilities, namely prompts \textbf{P17}--\textbf{P21} in Table \ref{tab:prompt_templates}, providing source code as prescribed in \cite{pearce2023examining}. Except for \textbf{P21}, the other prompts (\textbf{P17}--\textbf{P20}) do not include vulnerability information and often fail to generate reasonable patches, as reflected by their lower reasoning scores in Table~\ref{tab:existing_technique_real} for real-world samples in \dataset. Prompt \textbf{P21}, which uses the vulnerability description as the message, achieves a reasoning score similar to \textbf{P1}. This further proves that by failing to incorporate auxiliary information, the prompt templates proposed by Pearce et al. are unable to generate more reasonable patches for logical vulnerabilities.


\subsection{Effectiveness of Evaluation Metrics for Reasoning}\label{sec:appendix:effectiveness_eval_metrics}
\label{subsec:new_eval_metric_effectiveness}
In this experiment, we assess how much our automatically calculated reasoning metrics align with human assessments of patch quality and, in turn, their reliability. We randomly selected 200 off-the-shelf LLM-generated patches from our off-the-shelf evaluation and manually annotated them as \emph{reasonable} or \emph{unreasonable} using the criteria in Section \ref{subsubsec:reasoning}. Two security-knowledgeable annotators independently labeled each patch as reasonable or unreasonable, blinded to all automated scores. Disagreements were rare (<5\%) and resolved through discussion. We evaluate (i) compilation success, (ii) the LLM-as-a-judge binary verdict (J), (iii) cosine similarity between patch code embeddings computed with \textit{Microsoft/unixcoder-base}, (iv) ROUGE-L over patch, and (v) codeBLEU over patch. For each continuous metric, we treat patches above a percentile threshold as inferred reasonable, and report both the 70th and 90th percentile thresholds, including the corresponding threshold values. Table \ref{tab:reasoning_evaluation_scores} reports precision, recall, F1 score, and accuracy for all metrics.

\begin{table}[t]
\centering
\scriptsize
\setlength{\tabcolsep}{3pt}
\renewcommand{\arraystretch}{1.05}
\begin{tabular}{|p{0.34\columnwidth}|c|c|c|c|}
\hline
\textbf{Metric} & \textbf{Precision} & \textbf{Recall} & \textbf{F1 Score} & \textbf{Accuracy} \\
\hline
Compilation & 0.424 & 0.6 & 0.497 & 0.571 \\
\hline
J-Qwen & \textbf{0.773} & 0.667 & \textbf{0.716} & \textbf{0.8} \\
\hline
J-OpenAI & 0.5 & 0.721 & 0.59 & 0.701 \\
\hline
J-LLaMA & 0.547 & \textbf{0.839} & 0.662 & 0.677 \\
\hline
CodeEmbed 70th (0.944) & 0.47 & 0.508 & 0.488 & 0.622 \\
\hline
CodeEmbed 90th (0.981) & 0.615 & 0.262 & 0.368 & 0.68 \\
\hline
RougeL 70th (0.822) & 0.452 & 0.4 & 0.424 & 0.62 \\
\hline
RougeL 90th (0.957) & 0.682 & 0.214 & 0.326 & 0.69 \\
\hline
CodeBLEU 70th (0.708) & 0.352 & 0.357 & 0.355 & 0.545 \\
\hline
CodeBLEU 90th (0.906) & 0.579 & 0.157 & 0.247 & 0.665 \\
\hline
\end{tabular}
\caption{Precision, recall, F1 score, and accuracy across metrics.}
\label{tab:reasoning_evaluation_scores}
\end{table}

We observe that compilation is the weakest indicator of patch reasonableness, having the lowest accuracy (0.571). In contrast, the LLM-as-a-judge metric using Qwen achieves the highest accuracy (0.8). While several metrics achieve higher accuracy at the 90th percentile threshold, this is largely driven by class imbalance, as the dataset is naturally skewed toward the unreasonable patch class (130/200). Conservative thresholds tend to label most patches as unreasonable, thereby inflating accuracy. Overall, these results suggest that LLM-as-a-judge is a reasonably reliable metric that aligns most closely with human judgments, although there remains room for improvement.

\section{Conclusion}

We introduced \system, a systematic evaluation framework that leverages LLMs for patch validation. We also curated \dataset, a benchmark dataset of logical vulnerabilities. Using \system, we evaluated traditional AVR tools, LLM-based systems, and off-the-shelf LLMs across multiple dimensions and identified some key takeaways that would be helpful for future research.
\section*{Limitations}\label{sec:limitations}

We discuss some of the limitations of \system, and our constructed dataset \dataset below.

\noindent\textbf{\textit{Size of LogicDS.}} Our benchmark dataset, \dataset, consists of 61 logical vulnerabilities drawn from real-world open-source projects. We focus on real vulnerabilities to reflect the conditions developers actually face and to measure how well existing techniques perform outside of synthetic settings. For our work, since we require providing full source code for evaluation, end-to-end validation of patches through compilation and, when available, executable-based testing, we are forced to limit ourselves to open-source projects to construct \dataset. We curate samples by prioritizing (i) popular and widely used projects (e.g., OpenSSL, GnuTLS), (ii) logical vulnerabilities with direct security impact (e.g., privilege escalation, location tracking, DoS), and (iii) publicly available commits for both the vulnerable version and its patch, along with build instructions and, if available, testing scripts.

Constructing each sample requires substantial manual effort. For each selected project, we scan CVEs, identify issues that match our definition of logical vulnerabilities, confirm that a corresponding patch is publicly available, and then manually locate the core fix, as discussed in Section \ref{sec:appendix:locating_core_fix}. We additionally ensure the codebase compiles correctly (and run available tests when possible) so that each sample supports reliable compilation checking and practical evaluation. This high curation cost per sample make it challenging to scale \dataset to much larger sizes (e.g., thousands of samples).

\noindent\textbf{\textit{Assuming Perfect Localization and its Implications.}}  A real-world security vulnerability mitigation workflow consists of four primary stages: (i) vulnerability detection, (ii) root-cause analysis and localization, (iii) patch generation, and (iv) validation. Each stage introduces distinct technical challenges, and extensive prior work exists across these phases, including automated detection \cite{ullah2023can}, localization \cite{yu2024llm}, patch generation \cite{li2025sok}, and regression and security testing for validation \cite{zhou2024leveraging}. However, none of these works produces the expected output (i.e., 100\% sound and complete). As a result, all prior approaches focus on only one stage (e.g., localization) of the vulnerability mitigation workflow and assume a perfect solution for prior stages (e.g., vulnerability detection).

Consistent with prior work, this work aims to independently evaluate the \emph{patch generation} stage for logical vulnerabilities, isolating it from the effects of vulnerability detection and localization. Thus, we assume perfect vulnerability detection and localization, which can be performed manually or semi-automatically by first applying existing automated techniques and then refining them manually. Without this assumption, localization errors would propagate into the patch-generation stage, and failures could stem from incorrect/partial localization rather than limitations of the patch generator itself, making it difficult to identify the true strengths and weaknesses of repair tools. This is why we adopted perfect localization rather than an end-to-end testing approach that also involves error-prone automated localization.

\noindent\textbf{\textit{Adoption of samples from existing repair dataset.}} Existing dataset curation efforts \cite{bhandari2021cvefixes, mei2024arvo} collect CVEs and public commits based on CWE categories. However, several challenges limit their applicability for our purposes: (i) they do not explicitly target logical vulnerabilities, which often do not map cleanly to specific CWEs, (ii) they do not localize the patch in a commit, which often modifies several files at once where not all of the modifications are related to the vulnerability itself, (iii) these datasets also typically do not directly provide test scripts, which are often provided in commits separate from the patch itself, and (iv) most of their collected samples are memory-corruption vulnerabilities. Consequently, we adopted a manual curation approach to construct \dataset. However, evaluation result of samples in \dataset was still enough to confirm the patterns observed in our evaluations as several patches demonstrated such behavior to confirm our observation. We plan to elaborate \dataset in the future and also make the curation scripts and guidelines publicly available for future elaboration and research.

\noindent\textbf{\textit{Adoption of older CVEs.}} A common concern for generating patches for known CVEs is that LLMs might simply reproduce the fixes they saw during training. However, studies~\cite{jimenez2023swe} demonstrated that even if an LLM has encountered earlier versions of the code while training, it cannot memorize and regenerate those patches. Our patch validation analysis and manual observation also support this claim.

\noindent\textbf{\textit{Adopting ground-truth fix to reason patches.}} We measure reasoning by similarity to the ground-truth patch, which could undervalue correct fixes that implement the same logic differently. However, in our experiments, every plausible patch (i.e., a patch that passed both compilation and testing) scored above the $80^{th}$ percentile in cosine similarity and was judged similar to the ground truth by the LLM, so we observed no such misclassifications.

\noindent\textbf{\textit{Variations in reasoning scores among LLMs.}} We observe that the reasoning scores differ when we use different judging LLMs. For example, Qwen, as a judge, provides lower reasoning scores across all prompts compared to Llama and OpenAI. However, across different prompt approaches, the scores display the same trend, justifying our assumptions based on the reasoning scores.

\noindent\textbf{\textit{Evaluating more baselines and LLMs.}} We primarily focused on repair baselines that can potentially produce patches that are plausible. Several categories of AVR approaches e.g., template-guided \cite{shaw2014automatically, huang2019using}, constraint-based \cite{chida2022repairing,xuan2016nopol}, and several search-based approaches \cite{marginean2019sapfix} are inapplicable to apply for repairing logical vulnerabilities primarily due to focusing on specific templates, struggling with path explosion when analyzing source code, or only being able to reuse existing source code. A detailed analysis of several categories of vulnerability repair approaches and their limitations in repairing logical vulnerabilities is presented Appendix \ref{sec:appendix:characterizing_avr}.

\noindent\textbf{\textit{Circularity Risk of Overlapping LLM Judges.}} In our automated reasoning metrics, we used LLMs as judges for many of the metrics (e.g., LLM as binary judge \emph{J}). However, potentially overlapping LLMs for both patch generation and evaluation may introduce circularity. This can cause the same LLM to generate a patch and then evaluate its own patch suggestion, leading to bias \cite{panickssery2024llm,wataoka2024self}. To avoid this, we explicitly excluded any model from judging patches it generated. Concretely, a patch was evaluated only by Qwen3 and OpenAI models; a Qwen3-generated patch was judged only by Llama and OpenAI, and so on. Since patch generation and validation are performed independently, no LLM ever evaluates its own output in our experiments, mitigating the circularity concern.

\noindent\textbf{\textit{Manual efforts.}} We primarily require manual effort to (i) localize vulnerability and (ii) evaluate patches for reasoning. To reduce the manual effort of inspecting all prompts, we use reasoning metrics to identify patches that score significantly higher under one prompt than the other. To uncover common failure patterns, we manually inspect a fixed subset of patches with low scores. Prioritizing patches according to their reasoning metrics allows us to significantly reduce the number of manually reviewed patches.

\section*{Acknowledgements}
We thank the anonymous reviewers for their feedback and suggestions. Also, our use of any AI assistant during writing was limited to purely language-related assistance and short-form input. 

This work has been supported by the NSF under grants 2145631, 2215017, and 2442825, and the Public Wireless Supply Chain Innovation Fund (PWSCIF) under Federal Award ID Number 51-60-IF007.

\bibliography{ref}

\section*{Appendices}
\appendix
\sethlcolor{yellow!20}



\section{Characterizing AVR Approaches and Evaluation Frameworks for Logical Vulnerabilities} \label{sec:appendix:characterizing_avr}

\subsection{Automatic Vulnerability Repair (AVR)}\label{subsec:automatic_vulnerability_repair}




\begin{table*}[ht]
    \centering
    \begin{tabular}{|p{1.4cm}|p{4.8cm}|p{2.4cm}|p{5cm}|}
        \hline
        \textbf{Category} & \textbf{Optimal Scenario (When it Works)} & \textbf{Adaptable for Logical Vulnerability Repair} & \textbf{Limitations When Applying for Logical Vulnerabilities} \\
        \hline
        Template guided & When the vulnerability fix follows a template/pattern & No & Logical vulnerabilities does not follow any general fix pattern. \\
        \hline
        Constraint based & When the vulnerability can be fixed through enforcing constraints easily extractable from source code & Yes, but difficult & Constraint extraction for logical vulnerabilities struggle with path explosion in large codebases and often lacks the necessary specifications or testcases to derive required constraints. \\
        \hline
        Search based & When the vulnerability can be fixed by reusing or slightly adapting a similar code segment elsewhere in the source project & Yes, but achieves poor performance & Usually logical vulnerabilities cannot be fixed by reusing existing code, as each one has distinct expected behavior and requires its own custom logic. \\
        \hline
        Deep learning based & When the vulnerability fix follows a pattern learnable by deep learning models & Yes, but achieves poor performance & Usually logical vulnerabilities do not follow such recognizable pattern \\
        \hline
    \end{tabular}
    \caption{Analysis of existing AVR techniques and limitations towards repairing logical vulnerabilities}
    \label{tab:repair_approaches_summary}
\end{table*}

Automatic Vulnerability Repair (AVR) techniques aim to automatically fix software vulnerabilities. Their primary objective is to propose a patch that (i) eliminates the identified vulnerability and (ii) preserves the original functionality after the fix. In practice, vulnerability repair comprises three main stages: \emph{vulnerability localization}, \emph{patch generation}, and \emph{patch validation}, which mirror the steps developers naturally follow when repairing vulnerable code. 

In this work, evaluate the \emph{patch generation} phase of logical vulnerabilities. We identify the primary challenges for automatically generating fixes for such vulnerabilities, design frameworks to systematically evaluate the correctness of the generated fixes, and discuss the strengths and limitations of existing non-LLM and LLM approaches.

Existing AVR approaches primarily focus on repairing memory corruption vulnerabilities and can be broadly classified into two categories: \emph{non-learning-based} and \emph{learning-based} approaches. A recent study by Li et al.~\cite{li2025sok} provides a comprehensive analysis of various categories of AVR frameworks in addressing those vulnerabilities. In this section, we build upon previous studies by examining each category of AVR approaches when applied to logical vulnerability repair, highlighting the unique challenges and limitations they face in this context, which are 
summarized in Table \ref{tab:repair_approaches_summary}.



\subsubsection{Non-Learning Based Approaches}\label{subsubsec:nonlearning_based_approaches}
These non-machine learning based methods fall into three categories based on their repairing strategy: \textit{template-guided}, \textit{search-based}, and \textit{constraint-based} approaches.

\noindent\textbf{\textit{Template-guided approaches.}} Template-guided approaches rely on patterns of vulnerability properties \cite{shaw2014automatically, huang2019using} or historical patches \cite{xing2024if, zhang2022example} to generate patches. 
Due to their reliance on fixed patterns, these methods usually focus on vulnerabilities exhibiting recurring structures (e.g., buffer overflow \cite{shaw2014automatically}, null pointer dereference \cite{xing2024if}, API misuses \cite{zhang2022example}) where known fix templates can be reliably applied. In contrast, logical vulnerabilities typically involve highly context-specific behaviors and lack such common patterns, rendering these approaches inapplicable. 

\noindent\textbf{\textit{Constraint-based approaches.}} These methods extract program constraints that need to be satisfied to eliminate a vulnerability. These constraints are either extracted through (i) static analysis \cite{oh2018memfix, chida2022repairing}, (ii) symbolic execution \cite{shariffdeen2021concolic}, or (iii) dynamic analysis through running against a set of test suites~\cite{xuan2016nopol, agrawal1990dynamic}. 
(i) Among \textit{static analysis–based} techniques, Lee et al. \cite{oh2018memfix} leverage typestate checking to detect memory vulnerabilities. Liu et al. \cite{liu2023program} extract and modify Datalog facts to repair Java null-pointer exceptions, Python data leaks, and Solidity smart contract flaws, whereas Xu et al. \cite{xu2020automatic} build a mathematical model from historical patches and use automated reasoning to generate hotpatches.
However, these approaches rely on carefully defined properties to drive the static analysis and can only target specific vulnerability classes. Each logical vulnerability has its own distinct expected behavior, and formulating a corresponding property would require domain knowledge and significant manual effort comparable to writing the repair manually.
(ii) Symbolic execution–based methods derive the constraints needed to satisfy given testcases \cite{shariffdeen2021concolic, mechtaev2016angelix}. Consequently, they rely on a comprehensive, vulnerability-specific test suite to extract precise constraints. However, real-world logical vulnerabilities often provide only a single or no exploit trace, which is insufficient to recover the necessary constraints \cite{li2025sok}. 
(iii) Dynamic testing-based approaches \cite{xuan2016nopol, agrawal1990dynamic} also require a comprehensive test suite focusing on the vulnerability. Gao et al. \cite{gao2021beyond} eliminate this requirement by using address sanitizers to extract crash‐free constraints. However, logical vulnerabilities rarely cause program crashes and cannot be captured using address sanitizers, rendering this approach inapplicable.



\noindent\textbf{\textit{Search-based approaches.}} Search-based approaches explore a hypothetical search space for repair through mutations in existing code \cite{le2012systematic, marginean2019sapfix} or by extracting similar code from other functions or source files within a project \cite{jiang2018shaping, le2016history}. However, their ability to repair a vulnerability depends on the availability of similar code in their search base that can either be directly placed or adopted with minimal changes to fix a vulnerability. In contrast, it is usually challenging to find similar code that can be used to fix a real-world logical vulnerability, as each logical vulnerability has a distinct expected behavior and thus distinct logic that can be used to fix it. As a result, although we can technically apply search-based mechanisms to repair logical vulnerabilities, their performance is poor, which we demonstrate in this paper using SimFix \cite{jiang2018shaping} in \S\ref{subsec:baseline_performance}. 

\subsubsection{Learning Based Approaches}\label{subsubsec:learning_based_approaches}

Learning-based approaches pivot on machine learning techniques and can be broadly divided into two categories: deep learning-based and large language model (LLM)-based approaches. 

\noindent\textbf{\textit{Deep learning-based approaches.}} Deep learning-based approaches adopt deep learning models and treat the program repair problem as a neural machine translation (NMT) problem, providing the faulty code as input and obtaining the fixed code as output.
In this direction, SequenceR  \cite{chen2019sequencer} 
is the first to adopt an encoder-decoder based supervised recurrent neural network (RNN) machine translation model to generate patches. 
Later on, CURE \cite{jiang2021cure} improves on previous techniques on NMT-based program repair through subword tokenization and a code-aware token search strategy to provide more accurate patches. A recent work, KNOD \cite{jiang2023knod}, further increases the accuracy on Java program vulnerabilities through leveraging a novel tree-based decoder and emitting an abstract syntax tree (AST) to repair programs.

Although these approaches perform better than non-learning-based approaches \cite{lutellier2020coconut,jiang2023knod}
they require a dataset of vulnerabilities and their fixes to train the model. Furthermore, while training, the model attempts to extract patterns in source code that can be used to generate patches for specific vulnerabilities. Intuitively, these models work well when the underlying vulnerability can be fixed using a language safety pattern (e.g., bounds checking for buffer overflow vulnerabilities). In contrast, logical vulnerabilities do not follow any patterns. As we show in \S\ref{subsec:baseline_performance}, the state-of-the-art repair framework KNOD with a deep learning approach perform poor in fixing logical vulnerabilities. 



\subsubsection{Large Language Models for AVR}\label{subsubsec:llm_for_avr}

Recently, Large Language Models (LLMs) trained on vast training data have shown the ability to generate high-quality natural language outputs that are widely applicable across diverse areas such as text summarization \cite{jin2024comprehensive} and question answering \cite{saito2024unsupervised}. Due to similarities between code and natural language, researchers have explored LLMs in programming-related tasks such as code generation \cite{liu2024your} and analysis \cite{fang2024large}, and have been shown to be effective in understanding program syntax and semantics and analyzing program behaviors.

Several LLM-based approaches have been recently proposed to generate repair patches \cite{jin2023inferfix, kulsum2024case, pearce2023examining}. However, their capabilities and limitations are unknown for logical vulnerabilities. Also, LLMs have the potential to understand security invariants, even though they face limitations when computing complex constraints \cite{li2025sok}. As such, LLMs and LLM-based approaches appear promising for repairing logical vulnerabilities. However, to our knowledge, no prior research has systematically analyzed their capabilities and limitations in this specific context. 


\subsection{Limitations of AVR Evaluation Frameworks}\label{subsec:limitations_of_avr_eval_framework}

Among existing evaluations on LLMs for security vulnerabilities, Pearce et al. \cite{pearce2025asleep} assess the security of code generated by GitHub Copilot. Sandoval et al. \cite{sandoval2023lost} conducted a user study in which students used LLM assistance to write code, and the resulting programs were evaluated for security vulnerabilities. Deng et al. \cite{deng2024pentestgpt} propose and evaluate the use of LLMs for penetration testing. Ullah et al. \cite{ullah2023can} automatically evaluates LLMs with various prompts to identify security vulnerability in source codes. However, none of them particularly addresses repairing vulnerabilities. Jimenez et al. \cite{jimenez2023swe} evaluate how LLMs generate fixes for GitHub issues; however, most of these issues are not labeled to indicate whether they address security vulnerabilities. Among evaluation works that focus on automatically repairing security vulnerabilities, Li et al. \cite{li2025sok} systematically study existing AVR approaches, and provide a taxonomy, discussing pros and cons of each category of approaches. They also benchmark several existing works based on existing datasets. However, they do not focus on logical vulnerabilities and the limitations of using existing AVR approaches when applied to logical vulnerabilities.

Most related to our work is that of Pearce et al. \cite{pearce2023examining}, which evaluates the performance of LLMs towards repairing security vulnerabilities with zero-shot prompts. However, their evaluation framework have several limitations towards evaluating logical vulnerabilities: (i) Their evaluation approach do not take auxiliary information related to the vulnerability, such as vulnerability description, specification, or repair steps, etc., as input, which facilitates generation of reasonable patches for logical vulnerabilities (\S\ref{subsubsec:impact_adjusting_aux_information}), (ii) They lack any automated metrics for evaluation to indicate reasoning / identify reasonable patches. They only manually evaluate some patches of the Extractfix \cite{gao2021beyond} dataset and label them as identical, semantically equivalent, or reasonable, and (iii) They use address sanitizers and CodeQL \cite{codeql2021} to test patches, which are not suitable for logical vulnerabilities.

\section{Locating The Core Fix of a Patch}\label{sec:appendix:locating_core_fix}

\begin{lstlisting}[
    language=C,
    caption={Fix for CVE-2014-0224~\cite{henson2014cve2014-0224}.},
    label=lst:ssl3-ccs-diff,
    escapeinside={(*@}{@*)},
    float=ht,
    aboveskip=0cm,
    belowskip=-0.5cm
]
int ssl3_accept(SSL *s)
...
(*@\lstbg{green!20}{++   \hashchar define SSL3\_FLAGS\_CCS\_OK     0x0080;}@*)
...
    case SSL3_ST_SR_FINISHED_B:
(*@\lstbg{green!20}{++     s->s3->flags |= SSL3\_FLAGS\_CCS\_OK;}@*)
...

int ssl3_read_bytes(SSL *s, int type, unsigned char *buf, int len, int peek)
...
(*@\lstbg{green!20}{++~~~if (!(s->s3->flags \& SSL3\_FLAGS\_CCS\_OK))}@*)
(*@\lstbg{green!20}{++~~~\opencb }@*)
(*@\lstbg{green!20}{++~~~~~~al = SSL\_AD\_UNEXPECTED\_MESSAGE;}@*)
(*@\lstbg{green!20}{++~~~~~~SSLerr(SSL\_F\_SSL3\_READ\_BYTES,}@*)
(*@\lstbg{green!20}{~~~~~~~~~~~~~~~SSL\_R\_CCS\_RECEIVED\_EARLY);}@*)
(*@\lstbg{green!20}{++~~~~~~goto f\_err;}@*)
(*@\lstbg{green!20}{++~~~\closecb }@*)
(*@\lstbg{green!20}{++}@*)
(*@\lstbg{green!20}{++~~~s->s3->flags \&= ~SSL3\_FLAGS\_CCS\_OK;}@*)
...
}
\end{lstlisting}

We define the core fix as the portion of the patch that implements the decision logic needed to eliminate a vulnerability (e.g., the key authorization/validation/state-transition guard). As an example of a core fix, Listing \ref{lst:ssl3-ccs-diff} shows the fix for CVE-2014-0224 \cite{henson2014cve2014-0224}, where we observe a macro definition at Line 3, assignment of the macro to a variable at lines 6 and 19, and a check based on the variable and macro at lines 11--17. However, the primary fix is to add the check at lines 11--17, which enforces a check based on the flag and ensures that a ChangeCipherSpec message is received when expected. We refer to the \emph{if} block at lines 11--17 as the \emph{core fix}. In C/C++ and Java, the core fix typically corresponds to a compound statement enclosed in braces, with the function body as the top-level block.

During identification and annotation, two security-knowledgeable human annotators selected the hunk(s) they believed contained this decision logic. Disagreements during annotation were rare ($<5\%$) and were resolved through discussion to reach a consensus. We manually identify and localize the core fix by observing the vulnerable and fix commits, the vulnerability description $D$, and, if available, the behavioral specification $S$. 

During manual localization, we identify both the \emph{enclosing function} and the \emph{enclosing block}. The enclosing block provides fine-grained context for patch generation, while the enclosing function supplies the broader program context needed to understand the fix.
To localize the enclosing function, we extract its full body and signature. To localize the enclosing block, we identify the smallest brace-enclosed region that contains the core fix. If the block appears within a conditional or loop, we include the associated control structure (e.g., the condition). In general, this block may range from a small basic block to the entire function body.

When the fix logic is split across multiple hunks that are jointly required (e.g., a flag assigned in one place and enforced in another), in cases where possible, we expand the core-fix span to include all complementary hunks necessary to realize the intended security behavior, and treat that expanded region as the single-hunk core fix target for our evaluation. For example, in the fix presented in Listing \ref{lst:core-fix-multihunk-example} (representative of CVE-2023-0465, simplified for exposition), we place the entire block within the \texttt{for} loop as the core fix.

\begin{lstlisting}[
    language=C,
    caption={Simplified fix from CVE-2023-0465},
    label={lst:core-fix-multihunk-example},
    escapeinside={(*@}{@*)}
]
(*@{\setlength{\fboxsep}{1pt}\colorbox{red!15}{\strut\ttfamily\upshape\detokenize{- int i;}}}@*)
(*@{\setlength{\fboxsep}{1pt}\colorbox{green!15}{\strut\ttfamily\upshape\detokenize{+ int i, saw_err = 0;}}}@*)

// - - - start of hunk - - -
(*@{\setlength{\fboxsep}{1pt}\colorbox{red!15}{\strut\ttfamily\upshape\detokenize{- for (i = 1; i < N; i++) \{}}}@*)
(*@{\setlength{\fboxsep}{1pt}\colorbox{green!15}{\strut\ttfamily\upshape\detokenize{+ for (i = 0; i < N; i++) \{}}}@*)
item = chain[i];
(*@{\setlength{\fboxsep}{1pt}\colorbox{green!15}{\strut\ttfamily\upshape\detokenize{+ if (item.has_invalid_policy) saw_err = 1;}}}@*)
FAIL_IF(item.has_invalid_policy);
\}

(*@{\setlength{\fboxsep}{1pt}\colorbox{green!15}{\strut\ttfamily\upshape\detokenize{+ if (!saw_err) return internal_error();}}}@*)
// - - - end of hunk - - -
return OK;
\end{lstlisting}

In cases where two similar hunks in different code locations implement essentially the same decision logic for different modes/configurations, we select one of them as the representative fix, since both reflect the same underlying repair logic. Finally, if the main logic resides in a newly added helper function, we treat the helper function body as part of the core-fix span (i.e., include it within the single localized region).

We perform this localization on both the vulnerable source code $S$ and the fixed code $F$, resulting in the corresponding localized blocks and functions: $V_b$, $V_f$, $F_b$, and $F_f$. The repair framework receives $V_b$ or $V_f$ as input, while $F_b$ and $F_f$ are used solely for evaluation.

In some cases, such as the fix for CVE-2023-5363~\cite{cve2023-5363}, the patch involves inserting logic between two basic blocks within a much larger compound statement. Including the entire enclosing block in such cases would dilute the relevance of the context. To preserve focus, if the enclosing block exceeds a token threshold (2048 in our case), we instead extract a focused span centered on the patch location, including the immediate predecessor and successor basic blocks surrounding the patch location (e.g., lines 227--250 in the vulnerable code for CVE-2023-5363). This threshold was intended to avoid context dilution while staying within a practical prompt token budget for some LLMs. This ensures precise and focused block-level localization.
\section{Curation of \dataset}\label{sec:appendix:curation_of_dataset}

For dataset curation, we first search CVEs of the selected targets from the years 2010 to 2024, and based on the CVE description and publicly available fix, we determine whether the vulnerability is a logical vulnerability and has security implications. While selecting vulnerability samples, we exclude memory-corruption vulnerabilities (e.g., buffer overflow) and vulnerabilities that are more related to safe programming practices rather than a specific logical issue (e.g., SQL injection). Then, we manually searched the internet to obtain the developer-provided fix commit for the open-source implementation, by examining commit logs and open source vulnerability websites (e.g., \cite{bugzillaredhat, nvd_nist}). A breakdown of the open-source projects used to construct \dataset is given in \cite{logiceval}.

An example of the different portions of a sample is demonstrated in Table \ref{tab:sample_input_manual}.

\begin{table*}[ht]
\centering
\begin{tabular}{|p{0.2\linewidth}|p{0.15\linewidth}|p{0.5\linewidth}|}
\hline
\textbf{Input}          & 
\textbf{Mandatory}          & \textbf{Example}                                                 \\ \hline
Vulnerable Source Code   & \ding{52}  & Commit \cite{cve20191543_buggy} which is the parent commit of the fix in \cite{cve20191543_fix}  \\
\hline
Fixed Source Code  & \ding{52}  & Commit \cite{cve20191543_fix} available in national vulnerability database \cite{cve20191543_vul} \\
\hline
Vulnerability Description & \ding{52} & CVE description for CVE-2019-1543 found in \cite{cve20191543_vul}                         \\ \hline
Behavioral Specification & \ding{56}            & Extracted from RFC 7539 \cite{rfc_7539} Section 2.8                                 \\ \hline
Context Lines      & \ding{56}       & Lines 21-31 and 140-147 in $crypto/x509/x509\_vfy.c$ in \cite{cve20191543_buggy}                \\ \hline
Repair Description & \ding{56} & A text description generated by observing the fixed source code \cite{cve20191543_fix}. \\
\hline
Compilation Scripts & \ding{52} & Compilation script generated manually following implementation documents      \\ \hline
Testing Scripts & \ding{56} & Test found \cite{cve20191543_test} by manually checking the fix commit \cite{cve20191543_fix} \\ \hline
Faulty Function and block & \ding{52} & Lines 323-429 and 359-363 respectively in $crypto/x509/x509\_vfy.c$ in \cite{cve20191543_buggy_file} \\ \hline
Fixed Function and block & \ding{52}  & Lines 500-607 and 537-541 respectively in $crypto/x509/x509\_vfy.c$ in \cite{cve20191543_fix_file} \\ \hline

\end{tabular}

\caption{Example input and vulnerability localization (CVE-2019-1543~\cite{cve20191543_vul}) for \system. 
}

\label{tab:sample_input_manual}
\end{table*}

\subsection{Constructing Synthetic Java Samples.} 
Manually creating such samples would require significant time and efforts. 
Thus, to alleviate the manual work, we develop an LLM-assisted semi-automated approach for synthetic sample creation, which will also help future research works in creating synthetic Java examples from real-world logical vulnerabilities written in other programming languages.

In this approach, we first manually create one synthetic sample based on a real-world vulnerability by replicating the vulnerable and fixed functions $V_f$ \& $F_f$, with an intent to preserve original variable and function names where possible. We also write testcases that validate the created patch.
Once we create one such example, we leverage LLMs to create subsequent examples, by adopting a few-shot technique. We do that by providing the manually created sample, and prompting it to generate similar synthetic examples given a vulnerable real-world function and vulnerability description. In the subsequent prompt, we provide the generated vulnerable Java function and the real-world $V_b$ and $F_b$. Finally, we prompt the LLM to generate testcases for validation. We also assist the LLM  iteratively generate syntactically correct Java code by providing compilation logs in case of any errors. This few-shot technique assists the LLM in creating precise synthetic examples when provided with a real-world sample by clearly demonstrating the task at hand. Note that, this semi-automated technique only intended to alleviate the manual work. Although the generated samples needed to be manually verified after generation, this approach is significantly less time-consuming than creating synthetic samples from scratch.

\begin{table*}[ht]
\centering
\resizebox{\linewidth}{!}{
\begin{tabular}{|@{}p{0.08\linewidth}|p{0.88\linewidth}@{}|}
\hline
\textbf{ID} & \textbf{Description} \\ 
\hline
\textbf{P1-P3} & Temperature set to 0.2, 0.5, and 0.9, respectively \\
\hline
\textbf{P4} & Task-oriented prompt \\ 
\hline
\textbf{P5} & Zero-shot prompt (same as \textbf{P1}) \\
\hline
\textbf{P6} & Few-shot prompt including an example real-world vulnerability description, and vulnerable and fixed blocks\\
\hline
\textbf{P7} & Chain-of-thought prompting, first prompt obtains a repair suggestion $R_{suggestion}$, which is placed in the second prompt that asks for patch \\
\hline
\textbf{P8} & Same as \textbf{P7}, but $R_{suggestion}$ is not added to the second prompt \\
\hline
\textbf{P9} & Provides vulnerable block $V_b$ as input source code (same as \textbf{P1}) \\
\hline
\textbf{P10} & Provides entire vulnerable function $V_f$ as input source code \\
\hline
\textbf{P11} & Provides vulnerable block along with additional context lines $V_{context}$ \\
\hline
\textbf{P12} & Provides no auxiliary information, only the vulnerable block $V_b$ \\
\hline
\textbf{P13} & Provides the vulnerability description $D$ (same as \textbf{P1}) \\
\hline
\textbf{P14} & Provides the vulnerability description $D$ and the specification $S$ \\
\hline
\textbf{P15} & Provides repair steps $R$ and vulnerability description $D$ \\
\hline
\textbf{P16} & Provides the repair steps $R$ only, does not provide $D$ \\
\hline
\textbf{P17} & Obtained from \cite{pearce2023examining}, vulnerable code provided with no help (n.h.)\\
\hline
\textbf{P18} & Obtained from \cite{pearce2023examining}, deletes $V_f$ and adds a comment ``bugfix: fixed logical vulnerability'' (s.1).\\
\hline
\textbf{P19} & Obtained from \cite{pearce2023examining}, deletes $V_f$ and adds comment ``fixed logical vulnerability'' (s.2).\\
\hline
\textbf{P20} & Obtained from \cite{pearce2023examining}, after a ``// BUG: logical vulnerability'' comment, $V_b$ is shown in commented‐out form, immediately followed by a ``// FIXED:'' section. (c.)
\\
\hline
\textbf{P21} & Obtained from \cite{pearce2023examining}, first include a ``// BUG: logical vulnerability'' comment, then a ``// MESSAGE: $D$'' line. Next, show $V_b$ as commented‐out lines, and finally introduce the corrected code under a ``// FIXED VERSION:'' header. (c.m.)
\\
\hline
\end{tabular}}
\caption{Prompt Descriptions for Different Prompts Tested With Off-the-shelf LLMs. For prompts \textbf{P1} to \textbf{P16}, by default, the vulnerable block $V_b$ is provided as source code, }
\label{tab:prompt_templates}
\end{table*}

\section{Adopting Baselines for \system}\label{sec:appendix:adopt_baseline}

Existing work on program repair can be broadly categorized into three main paradigms: (i) non learning-based approaches, (ii) deep learning-based approaches, and (iii) large language model (LLM)-based approaches. To evaluate the capabilities and limitations of each, we select representative state-of-the-art methods from each category. We aim to analyze their capabilities in repairing logical vulnerabilities and derive insights that can be leveraged in future automated repair frameworks, focusing on them.

\noindent\textbf{Non Learning-Based Program Repair.}
We chose a state-of-the-art search-based method, SimFix~\cite{jiang2018shaping}, among non-learning-based approaches. Template-based methods focus on specific classes of bugs and are unsuitable due to the diverse nature of logical vulnerabilities, whereas constraint-based methods struggle with path explosion in large projects and are often inapplicable due to the lack of necessary specifications or test cases.

SimFix operates by mining AST-level modifications from a corpus of buggy programs and their corresponding patches. During inference, it analyzes the buggy project and attempts to replace faulty code snippets with similar code snippets from previously mined patches. Our experiments allow SimFix to generate up to 100 patches per bug. Also, since SimFix works with line-level localization, \system provides a manually located line within the core fix most suitable as the fault location. Furthermore, we allow SimFix to collect candidate patches from the largest project in Defects4J \cite{just2014defects4j}, Closure, providing a sufficiently large search space for collecting candidate patches.


\noindent\textbf{Deep Learning-Based Program Repair.}  
For deep learning-based repair, we evaluate \textit{KNOD} \cite{jiang2023knod}, a state-of-the-art deep learning-based framework, due to its superior performance compared to others  \cite{jiang2021cure, li2020dlfix, zhu2021syntax}, on standard benchmarks \cite{just2014defects4j, lin2017quixbugs}.
KNOD uses a three-stage tree decoder to generate repaired Java AST patches, and incorporates domain-specific rules to enhance the accuracy of AST node predictions. KNOD defaults to an output token limit of 512; therefore, we only provide the vulnerable block $V_b$ as input.

\noindent\textbf{LLM-Based Program Repair.}  
We consider \textit{VRPilot} \cite{kulsum2024case} as our baseline for LLM-based repair methods because of its robust performance and widely adoption in recent program repair evaluations \cite{li2025sok}. 
VRPilot leverages chain-of-thought (CoT) prompting first to elicit reasoning about a required fix and then generate candidate patches. It refines patch quality through iterative feedback using logs from compilation errors, functional test failures, and security tests until a plausible patch is identified or a predefined iteration budget is reached. Since we are testing for logical vulnerabilities, and through our experiments (\S\ref{subsubsec:impact_adjusting_aux_information}) we observed that providing $D$ significantly improves performance, we add $D$ in VRPilot's initial reasoning prompt.

In addition to baseline‐specific adaptations, when evaluating a suggested patch from an adapted baseline for reasoning, \system prompts the judge LLM with the vulnerability description $D$, the vulnerable block $V_b$, and the suggested patch to extract an explanation $E$ for the patch.

\section{Prompts Used for LLM-based Experiments}\label{sec:appendix:prompts_used_llm}

The prompts used in our experiments are summarized in Table \ref{tab:prompt_templates}.

As an example of a concrete prompt, we provide the template of prompt \textbf{P7}:

\begin{promptblock}
\textcolor{DotsGray}{...}

Here is a portion of a code:

<code>

(Vulnerable source code block or function)

</code>

Here is a description of the specification telling what to do:

<specification>

(Natural-text specification)

</specification>
Here is a description of the vulnerability:

<vulnerability>

(Vulnerability description)

</vulnerability>

Provide a repair for the mentioned code snippet to fix the vulnerability.

Provide repaired code exactly to replace the original buggy code mentioned between <code> and </code>.

Provide repaired code between <repair> and </repair> tags.

The code between <repair> and </repair> will be directly copied to replace the code between <code> and </code>.

\textcolor{DotsGray}{...}
\end{promptblock}

Here, we also test with prompts \textbf{P15} and \textbf{P16} with a repair description. From the results of the suggested patches of these prompts shown in Table \ref{tab:aux_info_real}, we observe that they have much higher scores in all metrics, indicating LLMs can generate more reasonable patches if concrete repair steps can be provided. We do not focus on the reasoning scores for these two prompts, since the repair steps of ground-truth fixes were already provided to the LLM under testing.

\section{Results of Experiments in Synthetic Examples in \dataset}\label{sec:appendix:synthetic_example_results}

In this Section, we present the results from the synthetic examples in \dataset for our experiments in Section \ref{sec:experiment}.

\begin{table}[ht]
    \centering
    \setlength{\tabcolsep}{0.8pt}
    \tiny
    \resizebox{\linewidth}{!}{
    \begin{tabular}{c|cccccc|cccccc|cccccc}
        \toprule
        \multicolumn{1}{c|}{} & \multicolumn{6}{c|}{\textbf{Llama-3.1}} & \multicolumn{6}{c|}{\textbf{Qwen-2.5}} & \multicolumn{6}{c}{\textbf{OpenAI-o3-mini}} \\
        \cmidrule(lr){2-7}\cmidrule(lr){8-13}\cmidrule(lr){14-19}
        \textbf{P} & \textbf{C} & \textbf{T} & \textbf{CSQ} & \textbf{CSO} & \textbf{JQ} & \textbf{JO} & \textbf{C} & \textbf{T} & \textbf{CSL} & \textbf{CSO} & \textbf{JL} & \textbf{JO} & \textbf{C} & \textbf{T} & \textbf{CSL} & \textbf{CSQ} & \textbf{JL} & \textbf{JQ} \\
        \midrule
        \textbf{P1} & \cellcolor[RGB]{205,236,198}0.27 & \cellcolor[RGB]{232,247,228}0.05 & \cellcolor[RGB]{53,158,83}0.81 & \cellcolor[RGB]{91,184,106}0.76 & \cellcolor[RGB]{232,247,228}0.12 & \cellcolor[RGB]{186,228,179}0.42 & \cellcolor[RGB]{222,242,216}0.17 & \cellcolor[RGB]{234,247,230}0.07 & \cellcolor[RGB]{61,168,90}0.80 & \cellcolor[RGB]{73,175,97}0.76 & \cellcolor[RGB]{173,222,166}0.48 & \cellcolor[RGB]{173,222,166}0.50 & \cellcolor[RGB]{192,230,185}0.32 & \cellcolor[RGB]{189,229,183}0.13 & \cellcolor[RGB]{57,162,86}0.78 & \cellcolor[RGB]{62,169,91}0.76 & \cellcolor[RGB]{117,197,119}0.68 & \cellcolor[RGB]{176,224,170}0.34 \\
        \textbf{P2} & \cellcolor[RGB]{210,238,204}0.25 & \cellcolor[RGB]{232,247,228}0.05 & \cellcolor[RGB]{59,166,89}0.80 & \cellcolor[RGB]{101,189,110}0.75 & \cellcolor[RGB]{232,247,228}0.12 & \cellcolor[RGB]{181,226,174}0.44 & \cellcolor[RGB]{217,241,211}0.19 & \cellcolor[RGB]{232,247,228}0.08 & \cellcolor[RGB]{72,175,96}0.79 & \cellcolor[RGB]{99,188,110}0.74 & \cellcolor[RGB]{175,223,169}0.47 & \cellcolor[RGB]{170,221,164}0.51 & -- & -- & -- & -- & -- & -- \\
        \midrule
        \textbf{P3} & \cellcolor[RGB]{231,246,226}0.17 & \cellcolor[RGB]{238,249,235}0.03 & \cellcolor[RGB]{67,172,94}0.79 & \cellcolor[RGB]{101,189,110}0.75 & \cellcolor[RGB]{237,249,234}0.08 & \cellcolor[RGB]{195,232,188}0.38 & \cellcolor[RGB]{226,244,221}0.15 & \cellcolor[RGB]{232,247,228}0.08 & \cellcolor[RGB]{83,180,102}0.78 & \cellcolor[RGB]{99,188,110}0.74 & \cellcolor[RGB]{175,223,169}0.47 & \cellcolor[RGB]{175,223,169}0.49 & -- & -- & -- & -- & -- & -- \\
        \textbf{P4} & \cellcolor[RGB]{215,240,209}0.23 & \cellcolor[RGB]{224,243,218}0.07 & \cellcolor[RGB]{53,158,83}0.81 & \cellcolor[RGB]{110,194,115}0.74 & \cellcolor[RGB]{237,249,234}0.08 & \cellcolor[RGB]{186,228,179}0.42 & \cellcolor[RGB]{204,235,197}0.25 & \cellcolor[RGB]{228,245,223}0.10 & \cellcolor[RGB]{96,186,108}0.77 & \cellcolor[RGB]{135,205,134}0.71 & \cellcolor[RGB]{170,221,164}0.49 & \cellcolor[RGB]{163,218,157}0.54 & \cellcolor[RGB]{192,230,185}0.32 & \cellcolor[RGB]{189,229,183}0.13 & \cellcolor[RGB]{46,151,78}0.80 & \cellcolor[RGB]{52,157,83}0.78 & \cellcolor[RGB]{119,198,121}0.67 & \cellcolor[RGB]{188,229,181}0.30 \\
        \midrule
        \textbf{P5} & \cellcolor[RGB]{205,236,198}0.27 & \cellcolor[RGB]{232,247,228}0.05 & \cellcolor[RGB]{46,151,78}0.82 & \cellcolor[RGB]{101,189,110}0.75 & \cellcolor[RGB]{237,249,234}0.08 & \cellcolor[RGB]{179,225,172}0.45 & \cellcolor[RGB]{222,242,216}0.17 & \cellcolor[RGB]{228,245,223}0.10 & \cellcolor[RGB]{61,168,90}0.80 & \cellcolor[RGB]{112,194,116}0.73 & \cellcolor[RGB]{151,212,147}0.57 & \cellcolor[RGB]{176,224,170}0.48 & \cellcolor[RGB]{192,230,185}0.32 & \cellcolor[RGB]{195,232,188}0.12 & \cellcolor[RGB]{62,169,91}0.77 & \cellcolor[RGB]{62,169,91}0.76 & \cellcolor[RGB]{134,205,133}0.62 & \cellcolor[RGB]{194,231,187}0.28 \\
        \textbf{P6} & \cellcolor[RGB]{210,238,204}0.25 & \cellcolor[RGB]{224,243,218}0.07 & \cellcolor[RGB]{59,166,89}0.80 & \cellcolor[RGB]{119,198,121}0.73 & \cellcolor[RGB]{232,247,228}0.12 & \cellcolor[RGB]{195,232,188}0.38 & \cellcolor[RGB]{197,232,190}0.28 & \cellcolor[RGB]{228,245,223}0.10 & \cellcolor[RGB]{72,175,96}0.79 & \cellcolor[RGB]{112,194,116}0.73 & \cellcolor[RGB]{163,218,157}0.52 & \cellcolor[RGB]{183,227,177}0.45 & \cellcolor[RGB]{175,223,169}0.37 & \cellcolor[RGB]{162,218,156}0.17 & \cellcolor[RGB]{247,252,245}0.47 & \cellcolor[RGB]{247,252,245}0.46 & \cellcolor[RGB]{212,239,206}0.28 & \cellcolor[RGB]{237,248,233}0.08 \\
        \textbf{P7} & \cellcolor[RGB]{243,251,240}0.10 & \cellcolor[RGB]{242,250,239}0.02 & \cellcolor[RGB]{131,203,130}0.73 & \cellcolor[RGB]{91,184,106}0.76 & \cellcolor[RGB]{241,250,238}0.05 & \cellcolor[RGB]{210,238,204}0.30 & \cellcolor[RGB]{222,242,216}0.17 & \cellcolor[RGB]{234,247,230}0.07 & \cellcolor[RGB]{132,204,131}0.74 & \cellcolor[RGB]{73,175,97}0.76 & \cellcolor[RGB]{201,234,194}0.35 & \cellcolor[RGB]{183,227,177}0.45 & \cellcolor[RGB]{212,239,206}0.25 & \cellcolor[RGB]{195,232,188}0.12 & \cellcolor[RGB]{94,186,107}0.73 & \cellcolor[RGB]{86,182,103}0.73 & \cellcolor[RGB]{131,203,130}0.63 & \cellcolor[RGB]{188,229,181}0.30 \\
        \textbf{P8} & \cellcolor[RGB]{139,207,137}0.45 & \cellcolor[RGB]{242,250,239}0.02 & \cellcolor[RGB]{201,234,194}0.65 & \cellcolor[RGB]{176,224,170}0.66 & \cellcolor[RGB]{247,252,245}0.00 & \cellcolor[RGB]{247,252,245}0.03 & \cellcolor[RGB]{128,202,128}0.50 & \cellcolor[RGB]{247,252,245}0.00 & \cellcolor[RGB]{244,251,241}0.60 & \cellcolor[RGB]{238,249,235}0.59 & \cellcolor[RGB]{247,252,245}0.02 & \cellcolor[RGB]{246,252,243}0.05 & \cellcolor[RGB]{138,207,136}0.47 & \cellcolor[RGB]{223,243,217}0.07 & \cellcolor[RGB]{179,225,172}0.62 & \cellcolor[RGB]{179,225,172}0.61 & \cellcolor[RGB]{233,247,229}0.15 & \cellcolor[RGB]{244,251,241}0.03 \\
        \bottomrule
        \end{tabular}
    }
    \caption{Compilation, testing, and reasoning score results for different LLM configurations for synthetic vulnerabilities in \dataset. Column meanings are the same as in Table \ref{tab:baseline_results}. ``P'' represents the prompt IDs. Suffixes ``L'', ``Q'', and ``O'' after each ``CS'' and ``J'' represent judging LLM Llama, Qwen, and OpenAI-o3-mini, respectively.}
    \label{tab:llm_config_synthetic}
\end{table}

\begin{table}[ht]
    \centering
    \setlength{\tabcolsep}{0.8pt}
    \tiny
    \resizebox{\linewidth}{!}{
    \begin{tabular}{c|cccccc|cccccc|cccccc}
        \toprule
        \multicolumn{1}{c|}{} & \multicolumn{6}{c|}{\textbf{Llama-3.1}} & \multicolumn{6}{c|}{\textbf{Qwen-2.5}} & \multicolumn{6}{c}{\textbf{OpenAI-o3-mini}} \\
        \cmidrule(lr){2-7}\cmidrule(lr){8-13}\cmidrule(lr){14-19}
        \textbf{P} & \textbf{C} & \textbf{T} & \textbf{CSQ} & \textbf{CSO} & \textbf{JQ} & \textbf{JO} & \textbf{C} & \textbf{T} & \textbf{CSL} & \textbf{CSO} & \textbf{JL} & \textbf{JO} & \textbf{C} & \textbf{T} & \textbf{CSL} & \textbf{CSQ} & \textbf{JL} & \textbf{JQ} \\
        \midrule
        \textbf{P9} & \cellcolor[RGB]{215,240,209}0.23 & \cellcolor[RGB]{232,247,228}0.05 & \cellcolor[RGB]{53,158,83}0.81 & \cellcolor[RGB]{101,189,110}0.75 & \cellcolor[RGB]{235,248,231}0.10 & \cellcolor[RGB]{195,232,188}0.38 & \cellcolor[RGB]{215,240,209}0.20 & \cellcolor[RGB]{234,247,230}0.07 & \cellcolor[RGB]{54,159,84}0.81 & \cellcolor[RGB]{86,182,103}0.75 & \cellcolor[RGB]{168,220,162}0.50 & \cellcolor[RGB]{161,217,155}0.55 & \cellcolor[RGB]{199,233,192}0.30 & \cellcolor[RGB]{195,232,188}0.12 & \cellcolor[RGB]{57,162,86}0.78 & \cellcolor[RGB]{62,169,91}0.76 & \cellcolor[RGB]{131,203,130}0.63 & \cellcolor[RGB]{197,232,190}0.27 \\
        \textbf{P10} & \cellcolor[RGB]{161,217,155}0.40 & \cellcolor[RGB]{224,243,218}0.07 & \cellcolor[RGB]{101,189,110}0.76 & \cellcolor[RGB]{101,189,110}0.75 & \cellcolor[RGB]{237,249,234}0.08 & \cellcolor[RGB]{201,234,194}0.35 & \cellcolor[RGB]{162,218,156}0.40 & \cellcolor[RGB]{206,236,199}0.17 & \cellcolor[RGB]{83,180,102}0.78 & \cellcolor[RGB]{112,194,116}0.73 & \cellcolor[RGB]{179,225,172}0.45 & \cellcolor[RGB]{168,220,162}0.52 & \cellcolor[RGB]{135,205,134}0.48 & \cellcolor[RGB]{122,199,123}0.22 & \cellcolor[RGB]{69,173,95}0.76 & \cellcolor[RGB]{77,177,99}0.74 & \cellcolor[RGB]{106,191,113}0.72 & \cellcolor[RGB]{188,229,181}0.30 \\
        \textbf{P11} & \cellcolor[RGB]{215,240,209}0.23 & \cellcolor[RGB]{219,241,213}0.08 & \cellcolor[RGB]{59,166,89}0.80 & \cellcolor[RGB]{91,184,106}0.76 & \cellcolor[RGB]{239,249,236}0.07 & \cellcolor[RGB]{186,228,179}0.42 & \cellcolor[RGB]{191,230,184}0.30 & \cellcolor[RGB]{219,241,213}0.13 & \cellcolor[RGB]{61,168,90}0.80 & \cellcolor[RGB]{86,182,103}0.75 & \cellcolor[RGB]{173,222,166}0.48 & \cellcolor[RGB]{165,219,159}0.53 & \cellcolor[RGB]{155,214,150}0.43 & \cellcolor[RGB]{155,214,150}0.18 & \cellcolor[RGB]{69,173,95}0.76 & \cellcolor[RGB]{69,173,95}0.75 & \cellcolor[RGB]{106,191,113}0.72 & \cellcolor[RGB]{188,229,181}0.30 \\
        \bottomrule
        \end{tabular}
    }
    \caption{Compilation, testing, and reasoning score results for varying portions of the source code provided for synthetic vulnerabilities in \dataset. Column meanings are the same as in Table \ref{tab:baseline_results} and \ref{tab:llm_config_real}.}
    \label{tab:source_code_synthetic}
\end{table}

\begin{table}[ht]
    \centering
    \setlength{\tabcolsep}{0.8pt}
    \tiny
    \resizebox{\linewidth}{!}{
    \begin{tabular}{c|cccccc|cccccc|cccccc}
        \toprule
        \multicolumn{1}{c|}{} & \multicolumn{6}{c|}{\textbf{Llama-3.1}} & \multicolumn{6}{c|}{\textbf{Qwen-2.5}} & \multicolumn{6}{c}{\textbf{OpenAI-o3-mini}} \\
        \cmidrule(lr){2-7}\cmidrule(lr){8-13}\cmidrule(lr){14-19}
        \textbf{P} & \textbf{C} & \textbf{T} & \textbf{CSQ} & \textbf{CSO} & \textbf{JQ} & \textbf{JO} & \textbf{C} & \textbf{T} & \textbf{CSL} & \textbf{CSO} & \textbf{JL} & \textbf{JO} & \textbf{C} & \textbf{T} & \textbf{CSL} & \textbf{CSQ} & \textbf{JL} & \textbf{JQ} \\
        \midrule
        \textbf{P12} & \cellcolor[RGB]{127,201,127}0.48 & \cellcolor[RGB]{242,250,239}0.02 & \cellcolor[RGB]{201,234,194}0.65 & \cellcolor[RGB]{204,235,197}0.62 & \cellcolor[RGB]{247,252,245}0.00 & \cellcolor[RGB]{247,252,245}0.03 & \cellcolor[RGB]{138,207,136}0.47 & \cellcolor[RGB]{247,252,245}0.00 & \cellcolor[RGB]{247,252,245}0.59 & \cellcolor[RGB]{247,252,245}0.57 & \cellcolor[RGB]{245,251,242}0.05 & \cellcolor[RGB]{247,252,245}0.03 & \cellcolor[RGB]{158,216,152}0.42 & \cellcolor[RGB]{232,246,227}0.05 & \cellcolor[RGB]{185,227,178}0.61 & \cellcolor[RGB]{185,227,178}0.60 & \cellcolor[RGB]{236,248,232}0.13 & \cellcolor[RGB]{247,252,245}0.00 \\
        \textbf{P13} & \cellcolor[RGB]{218,241,212}0.22 & \cellcolor[RGB]{224,243,218}0.07 & \cellcolor[RGB]{53,158,83}0.81 & \cellcolor[RGB]{101,189,110}0.75 & \cellcolor[RGB]{237,249,234}0.08 & \cellcolor[RGB]{197,232,190}0.37 & \cellcolor[RGB]{219,241,213}0.18 & \cellcolor[RGB]{238,249,235}0.05 & \cellcolor[RGB]{54,159,84}0.81 & \cellcolor[RGB]{86,182,103}0.75 & \cellcolor[RGB]{173,222,166}0.48 & \cellcolor[RGB]{176,224,170}0.48 & \cellcolor[RGB]{192,230,185}0.32 & \cellcolor[RGB]{195,232,188}0.12 & \cellcolor[RGB]{62,169,91}0.77 & \cellcolor[RGB]{69,173,95}0.75 & \cellcolor[RGB]{131,203,130}0.63 & \cellcolor[RGB]{202,234,195}0.25 \\
        \textbf{P14} & \cellcolor[RGB]{240,249,236}0.12 & \cellcolor[RGB]{232,247,228}0.05 & \cellcolor[RGB]{89,183,105}0.77 & \cellcolor[RGB]{72,175,96}0.78 & \cellcolor[RGB]{224,243,218}0.17 & \cellcolor[RGB]{163,218,157}0.52 & \cellcolor[RGB]{230,246,225}0.13 & \cellcolor[RGB]{238,249,235}0.05 & \cellcolor[RGB]{61,168,90}0.80 & \cellcolor[RGB]{73,175,97}0.76 & \cellcolor[RGB]{121,199,122}0.68 & \cellcolor[RGB]{135,205,134}0.65 & \cellcolor[RGB]{232,246,227}0.17 & \cellcolor[RGB]{223,243,217}0.07 & \cellcolor[RGB]{57,162,86}0.78 & \cellcolor[RGB]{62,169,91}0.76 & \cellcolor[RGB]{96,186,108}0.75 & \cellcolor[RGB]{188,229,181}0.30 \\
        \textbf{P15} & \cellcolor[RGB]{148,211,144}0.43 & \cellcolor[RGB]{46,151,78}0.33 & \cellcolor[RGB]{67,172,94}0.79 & \cellcolor[RGB]{46,151,78}0.82 & \cellcolor[RGB]{61,168,90}0.72 & \cellcolor[RGB]{46,151,78}0.97 & \cellcolor[RGB]{99,188,110}0.58 & \cellcolor[RGB]{46,151,78}0.53 & \cellcolor[RGB]{61,168,90}0.80 & \cellcolor[RGB]{54,159,84}0.78 & \cellcolor[RGB]{46,151,78}0.97 & \cellcolor[RGB]{50,155,81}0.98 & \cellcolor[RGB]{173,222,166}0.38 & \cellcolor[RGB]{46,151,78}0.32 & \cellcolor[RGB]{62,169,91}0.77 & \cellcolor[RGB]{69,173,95}0.75 & \cellcolor[RGB]{46,151,78}0.95 & \cellcolor[RGB]{46,151,78}0.73 \\
        \textbf{P16} & \cellcolor[RGB]{152,213,148}0.42 & \cellcolor[RGB]{77,177,99}0.28 & \cellcolor[RGB]{89,183,105}0.77 & \cellcolor[RGB]{46,151,78}0.82 & \cellcolor[RGB]{46,151,78}0.79 & \cellcolor[RGB]{54,159,84}0.93 & \cellcolor[RGB]{118,197,120}0.53 & \cellcolor[RGB]{59,166,89}0.49 & \cellcolor[RGB]{96,186,108}0.77 & \cellcolor[RGB]{46,151,78}0.79 & \cellcolor[RGB]{54,159,84}0.93 & \cellcolor[RGB]{46,151,78}1.00 & \cellcolor[RGB]{189,229,183}0.33 & \cellcolor[RGB]{88,182,104}0.26 & \cellcolor[RGB]{77,177,99}0.75 & \cellcolor[RGB]{86,182,103}0.73 & \cellcolor[RGB]{46,151,78}0.95 & \cellcolor[RGB]{73,175,97}0.63 \\
        \bottomrule
        \end{tabular}
    }
    \caption{Compilation, testing, and reasoning scores for varying auxiliary information provided in prompts for synthetic vulnerabilities in \dataset. Column meanings are the same as in Table \ref{tab:baseline_results} and \ref{tab:llm_config_real}.}
    \label{tab:aux_info_synthetic}
\end{table}

\section{Statistical Significance Analysis of Primary Claims}\label{sec:appendix:statistical_significance}

\begin{table*}[ht]
\centering
\footnotesize
\setlength{\tabcolsep}{3pt}
\renewcommand{\arraystretch}{1.02}
\begin{tabularx}{\textwidth}{|c|>{\RaggedRight\arraybackslash}p{0.21\textwidth}|>{\RaggedRight\arraybackslash}p{0.20\textwidth}|>{\RaggedRight\arraybackslash}X|>{\RaggedRight\arraybackslash}p{0.19\textwidth}|}
\hline
\textbf{ID} & \textbf{Claim} & \textbf{C (McNemar)} & \textbf{CS (Wilcoxon p, Cliff $\delta$, Mean Diff., CI)} & \textbf{J (McNemar)} \\
\hline
1 & \textbf{P1} and \textbf{P2} show similar results & 86/219 vs 85/219 ($p$=1.0000, OR=1.14) & 0.7801 vs 0.7740; $p$=0.1516 (\textit{ns}); $\delta$=+0.080; $\Delta$=+0.0061 [-0.0018, +0.0140] & 169/438 vs 175/438 ($p$=0.5446, OR=0.84) \\
\hline
2 & \textbf{P1} and \textbf{P4} show similar results & 136/332 vs 141/332 ($p$=0.5114, OR=0.76) & 0.7753 vs 0.7696; $p$=0.1634 (\textit{ns}); $\delta$=+0.062; $\Delta$=+0.0057 [-0.0028, +0.0141] & 289/664 vs 283/664 ($p$=0.6137, OR=1.13) \\
\hline
3 & \textbf{P5} shows better \textbf{C}, \textbf{CS}, \textbf{J} than \textbf{P7} & 139/339 vs 114/339 ($p$=0.0015, OR=2.47) & 0.7757 vs 0.7420; $p \leq 0.001$ (\textit{sig}); $\delta$=+0.537; $\Delta$=+0.0337 [+0.0260, +0.0413] & 300/726 vs 285/726 ($p$=0.2513, OR=1.22) \\
\hline
4 & \textbf{P8} shows more \textbf{C} but less \textbf{CS}, \textbf{J} than \textbf{P7} & 196/339 vs 114/339 ($p \leq 0.001$, OR=5.56) & 0.6344 vs 0.7420; $p \leq 0.001$ (\textit{sig}); $\delta$=-0.884; $\Delta$=-0.1076 [-0.1218, -0.0939] & 54/726 vs 285/726 ($p \leq 0.001$, OR=0.08) \\
\hline
5a & \textbf{P12} has higher \textbf{C} but lower \textbf{CS}, \textbf{J} than \textbf{P13} & 189/339 vs 138/339 ($p \leq 0.001$, OR=3.12) & 0.6206 vs 0.7731; $p \leq 0.001$ (\textit{sig}); $\delta$=-0.967; $\Delta$=-0.1525 [-0.1662, -0.1389] & 53/726 vs 301/726 ($p \leq 0.001$, OR=0.04) \\
\hline
5b & \textbf{P12} has higher \textbf{C} but lower \textbf{CS}, \textbf{J} than \textbf{P14} & 189/339 vs 107/339 ($p \leq 0.001$, OR=6.47) & 0.6206 vs 0.7769; $p \leq 0.001$ (\textit{sig}); $\delta$=-0.950; $\Delta$=-0.1564 [-0.1715, -0.1411] & 53/726 vs 375/726 ($p \leq 0.001$, OR=0.04) \\
\hline
6 & \textbf{P11} has slightly better \textbf{C} than \textbf{P9} & 153/339 vs 138/339 ($p$=0.0444, OR=1.88) & 0.7758 vs 0.7789; $p$=0.6332 (\textit{ns}); $\delta$=+0.058; $\Delta$=-0.0031 [-0.0093, +0.0032] & 309/726 vs 301/726 ($p$=0.5264, OR=1.14) \\
\hline
7 & \textbf{P9} has lower \textbf{C} but higher \textbf{CS}, \textbf{J} than \textbf{P10} & 136/333 vs 174/333 ($p \leq 0.001$, OR=0.36) & 0.7784 vs 0.7579; $p \leq 0.001$ (\textit{sig}); $\delta$=+0.345; $\Delta$=+0.0205 [+0.0127, +0.0284] & 296/714 vs 289/714 ($p$=0.6110, OR=1.11) \\
\hline
8a & \textbf{P17} has lower \textbf{C}, \textbf{CS}, \textbf{J} than \textbf{P5} & 85/339 vs 139/339 ($p \leq 0.001$, OR=0.31) & 0.5683 vs 0.7757; $p \leq 0.001$ (\textit{sig}); $\delta$=-0.967; $\Delta$=-0.2074 [-0.2250, -0.1900] & 21/726 vs 300/726 ($p \leq 0.001$, OR=0.04) \\
\hline
8b & \textbf{P20} has lower \textbf{C}, \textbf{CS}, \textbf{J} than \textbf{P5} & 70/339 vs 139/339 ($p \leq 0.001$, OR=0.17) & 0.6656 vs 0.7757; $p \leq 0.001$ (\textit{sig}); $\delta$=-0.934; $\Delta$=-0.1100 [-0.1233, -0.0974] & 143/726 vs 300/726 ($p \leq 0.001$, OR=0.27) \\
\hline
\end{tabularx}
\caption{Paired significance test results for the primary claims, aggregated across all evaluated LLMs for each prompt type.}
\label{tab:statistical_significance_results}
\end{table*}

\begin{table}[ht]
    \centering
    \setlength{\tabcolsep}{0.8pt}
    \tiny
    \resizebox{\linewidth}{!}{
    \begin{tabular}{c|cccccc|cccccc|cccccc}
        \toprule
        \multicolumn{1}{c|}{} & \multicolumn{6}{c|}{\textbf{Llama-3.1}} & \multicolumn{6}{c|}{\textbf{Qwen-2.5}} & \multicolumn{6}{c}{\textbf{OpenAI-o3-mini}} \\
        \cmidrule(lr){2-7}\cmidrule(lr){8-13}\cmidrule(lr){14-19}
        \textbf{P} & \textbf{C} & \textbf{T} & \textbf{CSQ} & \textbf{CSO} & \textbf{JQ} & \textbf{JO} & \textbf{C} & \textbf{T} & \textbf{CSL} & \textbf{CSO} & \textbf{JL} & \textbf{JO} & \textbf{C} & \textbf{T} & \textbf{CSL} & \textbf{CSQ} & \textbf{JL} & \textbf{JQ} \\
        \midrule
        \textbf{P17} & \cellcolor[RGB]{229,245,224}0.18 & \cellcolor[RGB]{247,252,245}0.00 & \cellcolor[RGB]{240,250,237}0.58 & \cellcolor[RGB]{238,249,234}0.55 & \cellcolor[RGB]{247,252,245}0.00 & \cellcolor[RGB]{247,252,245}0.02 & \cellcolor[RGB]{247,252,245}0.00 & \cellcolor[RGB]{247,252,245}0.00 & \cellcolor[RGB]{247,252,245}0.59 & \cellcolor[RGB]{247,252,245}0.57 & \cellcolor[RGB]{245,251,242}0.05 & \cellcolor[RGB]{244,251,241}0.07 & \cellcolor[RGB]{230,246,226}0.18 & \cellcolor[RGB]{241,250,238}0.02 & \cellcolor[RGB]{204,235,197}0.58 & \cellcolor[RGB]{204,235,197}0.57 & \cellcolor[RGB]{247,252,245}0.02 & \cellcolor[RGB]{247,252,245}0.00 \\
        \textbf{P18} & \cellcolor[RGB]{243,251,240}0.10 & \cellcolor[RGB]{247,252,245}0.00 & \cellcolor[RGB]{221,242,215}0.62 & \cellcolor[RGB]{209,237,203}0.61 & \cellcolor[RGB]{247,252,245}0.00 & \cellcolor[RGB]{245,251,242}0.05 & \cellcolor[RGB]{247,252,245}0.00 & \cellcolor[RGB]{247,252,245}0.00 & \cellcolor[RGB]{244,251,241}0.60 & \cellcolor[RGB]{247,252,245}0.57 & \cellcolor[RGB]{247,252,245}0.03 & \cellcolor[RGB]{246,252,243}0.05 & \cellcolor[RGB]{246,252,244}0.08 & \cellcolor[RGB]{247,252,245}0.00 & \cellcolor[RGB]{198,233,191}0.59 & \cellcolor[RGB]{198,233,191}0.58 & \cellcolor[RGB]{242,251,240}0.07 & \cellcolor[RGB]{244,251,241}0.03 \\
        \textbf{P19} & \cellcolor[RGB]{229,245,224}0.18 & \cellcolor[RGB]{247,252,245}0.00 & \cellcolor[RGB]{232,247,228}0.60 & \cellcolor[RGB]{209,237,203}0.61 & \cellcolor[RGB]{247,252,245}0.00 & \cellcolor[RGB]{247,252,245}0.03 & \cellcolor[RGB]{247,252,245}0.00 & \cellcolor[RGB]{247,252,245}0.00 & \cellcolor[RGB]{234,247,230}0.62 & \cellcolor[RGB]{229,245,224}0.61 & \cellcolor[RGB]{242,251,240}0.07 & \cellcolor[RGB]{242,251,240}0.08 & \cellcolor[RGB]{243,251,240}0.10 & \cellcolor[RGB]{241,250,238}0.02 & \cellcolor[RGB]{192,230,185}0.60 & \cellcolor[RGB]{192,230,185}0.59 & \cellcolor[RGB]{241,250,238}0.08 & \cellcolor[RGB]{245,252,243}0.02 \\
        \textbf{P20} & \cellcolor[RGB]{243,251,240}0.10 & \cellcolor[RGB]{247,252,245}0.00 & \cellcolor[RGB]{169,221,163}0.69 & \cellcolor[RGB]{176,224,170}0.66 & \cellcolor[RGB]{239,249,236}0.07 & \cellcolor[RGB]{219,241,213}0.25 & \cellcolor[RGB]{245,252,243}0.02 & \cellcolor[RGB]{247,252,245}0.00 & \cellcolor[RGB]{216,240,210}0.65 & \cellcolor[RGB]{222,242,216}0.62 & \cellcolor[RGB]{230,246,226}0.18 & \cellcolor[RGB]{222,242,216}0.25 & \cellcolor[RGB]{246,252,244}0.08 & \cellcolor[RGB]{241,250,238}0.02 & \cellcolor[RGB]{135,205,134}0.68 & \cellcolor[RGB]{128,202,128}0.68 & \cellcolor[RGB]{193,231,186}0.38 & \cellcolor[RGB]{230,246,225}0.13 \\
        \textbf{P21} & \cellcolor[RGB]{247,252,245}0.07 & \cellcolor[RGB]{242,250,239}0.02 & \cellcolor[RGB]{89,183,105}0.77 & \cellcolor[RGB]{110,194,115}0.74 & \cellcolor[RGB]{222,242,216}0.18 & \cellcolor[RGB]{186,228,179}0.42 & \cellcolor[RGB]{247,252,245}0.00 & \cellcolor[RGB]{247,252,245}0.00 & \cellcolor[RGB]{132,204,131}0.74 & \cellcolor[RGB]{148,211,144}0.70 & \cellcolor[RGB]{142,209,140}0.60 & \cellcolor[RGB]{156,215,151}0.57 & \cellcolor[RGB]{247,252,245}0.07 & \cellcolor[RGB]{247,252,245}0.00 & \cellcolor[RGB]{69,173,95}0.76 & \cellcolor[RGB]{69,173,95}0.75 & \cellcolor[RGB]{102,190,111}0.73 & \cellcolor[RGB]{207,236,200}0.23 \\
        \bottomrule
        \end{tabular}
    }
    \caption{Compilation, testing, and reasoning score results using templates provided in \cite{pearce2023examining} for synthetic vulnerabilities in \dataset. Column meanings are the same as in Table \ref{tab:baseline_results} and \ref{tab:llm_config_real}.}
    \label{tab:existing_technique_synthetic}
\end{table}


To evaluate whether the observed claims in our Experiments Section (\S\ref{sec:experiment}) are statistically significant, we performed paired significance tests for each claim. For binary outcomes, compilation success/failure (\textbf{C}), test-case success/failure (\textbf{T}), and the binary LLM-as-a-judge verdict (\textbf{J}): we used an exact McNemar's test and report the discordant counts, p-value, and odds ratio (OR). For continuous outcomes, such as the LLM-as-a-judge cosine similarity score (CS) we used the Wilcoxon signed-rank test and report the p-value, Cliff's $\delta$ (effect size), the mean paired difference, and the bootstrap confidence interval (CI).

Our primary claims are as follows:
\begin{enumerate}
    \item Adjusting temperature has minimal impact on repair performance (i.e., compilation, testing and reasoning scores for output patches from prompts \textbf{P1}--\textbf{P3} are identical). We test this by comparing \textbf{P1} vs.\ \textbf{P2}.
    \item Orientation has minimal impact (i.e., output patches from prompt templates \textbf{P1} and \textbf{P4} yield similar results).
    \item The zero-shot prompt template (\textbf{P5}) achieves higher compilation success and slightly better reasoning than the CoT prompt template (\textbf{P7}).
    \item Including reasoning text in the patch-generation prompt improves patch quality. Specifically, \textbf{P8} increases compilation success but lowers reasoning scores relative to \textbf{P7}, which includes vulnerability text.
    \item Auxiliary information is essential for generating reasonable patches for logical vulnerabilities. Specifically, \textbf{P12} (no auxiliary info) performs worse than \textbf{P13} (vulnerability text) and \textbf{P14} (specification info), even though \textbf{P12} produces patches with a higher compilation success rate.
    \item Adding context (\textbf{P11}) to a vulnerable source-code block (\textbf{P9}) improves compilation, while reasoning scores remain similar.
    \item Providing only the vulnerable block (\textbf{P9}) yields more compilable patches and slightly better reasoning than providing the entire function (\textbf{P10}).
    \item Prompts from Pearce et al.\ (\textbf{P17}--\textbf{P20}) are less effective than our baseline prompt configuration (\textbf{P5}). We demonstrate this by comparing \textbf{P17} vs.\ \textbf{P5} and \textbf{P20} vs.\ \textbf{P5}.
\end{enumerate}

For each claim, we report results aggregated across all evaluated LLMs for each prompt type. The statistical significance results are provided in Table \ref{tab:statistical_significance_results}. The significance tests largely support all of our claims, with two minor exceptions. For the binary LLM-as-a-judge reasoning metric (\textbf{J}), the differences are not statistically significant for the \textbf{P5} vs. \textbf{P7} comparison in claim 3 and for the \textbf{P9} vs. \textbf{P10} comparison in claim 6. In both cases, however, the continuous reasoning metric (\textbf{CS}) remains statistically significant. In essence, the relative performance of \textbf{P9} against \textbf{P10} depends on the requirements for vulnerability repair. Vulnerabilities that require broader, function-level context benefit from being provided in \textbf{P10}, whereas issues that are localized to a specific region are often better addressed with a block-level snippet in \textbf{P9}, which can facilitate repair while avoiding incorrect localization.

\section{Example Codes Supporting Manual Inspection of Patches}\label{sec:appendix:example_codes}



\begin{table}[ht]
    \caption{Llama Fix for CVE-2015-1793 \cite{cve-2015-1793}}
    \label{tab:listing_1}
    \scriptsize
    \setlength{\tabcolsep}{1pt}
    \begin{tabular}{|p{\linewidth-6pt}|}
    \hline
    Before processing the certificate chain, we need to check if the certificate has a \hl{Basic Constraints extension} and if the cA flag is set. We can use the `X509\_get\_extension\_flags' function to \hl{check if the extension is critical} and the `X509v3\_get\_ext\_d2i' function to parse the extension. \\
    \hline
    \lstbg{codegray!20}{Suggested Fix} \\
\begin{lstlisting}[
        language=C,
        escapeinside={(*@}{@*)},
        aboveskip=-0.15cm,
        belowskip=-0.3cm
    ]
// Check the Basic Constraints extension
X509_EXTENSION *ext = X509_get_ext(xtmp, NID_basic_constraints);
if (ext!= NULL) {
  // Check if the extension is critical
  if(X509_get_extension_flags(ext) & (*@\lstbgt{red!20}{EXFLAG\_CRITICAL}@*)){
    ...
    if (bc!= NULL) {
      ...
      (*@\lstbgt{red!20}{BASIC\_CONSTRAINTS\_free(bc)}@*);
  ...
\end{lstlisting} \\
    \hline
    \end{tabular}

    \vspace{1mm}
    \small
    Based on the prompt (highlighted in yellow), 2 undefined variables (highlighted in red) are generated. 
\end{table}

\begin{table}[ht]
    \caption{Qwen Fix for CVE-2017-3142 \cite{cve-2017-3142}}
    \label{tab:listing_2}
    \scriptsize
    \setlength{\tabcolsep}{1pt}
    \begin{tabular}{|p{\linewidth-6pt}|}
    \hline
    \lstbg{codegray!20}{Suggested Fix} \\
\begin{lstlisting}[
        language=C,
        escapeinside={(*@}{@*)},
        aboveskip=-0.15cm,
        belowskip=-0.3cm
    ]
if (msg->tsigstatus != dns_rcode_noerror) {
} else if (tsig.error != dns_rcode_noerror) {
} else {
  // Extra check to ensure the requestor is authorized
  if ((*@\lstbgt{red!20}{!is\_requestor\_authorized(msg)}@*)) {
    result = (*@\lstbgt{red!20}{DNS\_R\_UNAUTHORIZED}@*);
  ...
\end{lstlisting} \\
    \hline
    \lstbg{codegray!20}{Ground Truth Fix} \\
\begin{lstlisting}[
        language=C,
        escapeinside={(*@}{@*)},
        aboveskip=-0.15cm,
        belowskip=-0.3cm
    ]
if (msg->verified_sig && msg->tsigstatus == dns_rcode_noerror && tsig.error == dns_rcode_noerror) {
} else if ((!msg->verified_sig) || (msg->tsigstatus != dns_rcode_noerror)) {
} else {
  INSIST(tsig.error != dns_rcode_noerror);
  result = DNS_R_TSIGERRORSET;
}
\end{lstlisting} \\
    \hline
    \end{tabular}

    \vspace{1mm}
    \small
    Undefined variables (highlighted in red) were created due to the lack of context. 
\end{table}

\begin{table}[ht]
    \caption{Llama Fix for CVE-2015-1793 \cite{cve-2015-1793}}
    \label{tab:listing_3}
    \scriptsize
    \setlength{\tabcolsep}{1pt}
    \begin{tabular}{|p{\linewidth-6pt}|}
    \hline
    The whole function was input in the prompt, but the LLM failed to locate the vulnerable portion and add extra code after the vulnerable portion instead. The original vulnerable code has 319 lines, and the suggested patch has 336 lines. \\
    \hline
    \lstbg{codegray!20}{Suggested Fix} \\
\begin{lstlisting}[
        language=C,
        escapeinside={(*@}{@*)},
        aboveskip=-0.15cm,
        belowskip=-0.3cm
    ]
if (ok > 0) {
  X509_free(xtmp);
  while (num > j) {
    xtmp = sk_X509_pop(ctx->chain);
    X509_free(xtmp);
    num--;
    ctx->last_untrusted--;
  }
  retry = 1;
  break;
}
...
end:
  ...
(*@\lstbg{red!20}{// Fix: Check for Basic Constraints cA value}@*)
(*@\lstbg{red!20}{if (ctx->chain != NULL) \opencb}@*)
(*@\lstbg{red!20}{~~for (i = 0; i < sk\_X509\_num(ctx->chain); i++) \opencb}@*)
(*@\lstbg{red!20}{~~~~x = sk\_X509\_value(ctx->chain, i);}@*)
(*@\lstbg{red!20}{~~~~if (X509\_check\_ca(x)!= X509\_CHECK\_CA\_OK) \opencb}@*)
(*@\lstbg{red!20}{~~~~~~ctx->error = X509\_V\_ERR\_INVALID\_CA;}@*)
(*@\lstbg{red!20}{~~~~~~~~ctx->current\_cert = x;}@*)
(*@\lstbg{red!20}{~~~~~~~~ctx->error\_depth = i;}@*)
(*@\lstbg{red!20}{~~~~~~~~ok = cb(0, ctx);}@*)
(*@\lstbg{red!20}{~~~~~~~~if (!ok)}@*)
(*@\lstbg{red!20}{~~~~~~~~~~goto end;}@*)
  ...
\end{lstlisting} \\
    \hline
    \lstbg{codegray!20}{Ground Truth Fix} \\
\begin{lstlisting}[
        language=C,
        escapeinside={(*@}{@*)},
        aboveskip=-0.15cm,
        belowskip=-0.3cm
    ]
if (ok > 0) {
  X509_free(xtmp);
  while (num > j) {
    xtmp = sk_X509_pop(ctx->chain);
    X509_free(xtmp);
    num--;
  }
  ctx->last_untrusted = sk_X509_num(ctx->chain);
  retry = 1;
  break;
}
\end{lstlisting} \\
    \hline
    \end{tabular}

    \vspace{1mm}
    \small
     
\end{table}

\begin{table}[ht]
    \caption{OpenAI Fix for CVE-2022-1434 \cite{cve_2022_1434}}
    \label{tab:listing_4}
    \scriptsize
    \setlength{\tabcolsep}{1pt}
    \begin{tabular}{|p{\linewidth-6pt}|}
    \hline
    The fix adds a check after calling OSSL\_PARAM\_locate\_const to verify that a valid pointer was returned rather than a NULL reference. Without this check, subsequent code that uses `p' could \hl{try to dereference a NULL pointer} if OSSL\_PARAM\_locate\_const doesn't find the expected parameter, potentially leading to undefined behavior, a crash, or even exploitable security vulnerabilities.\\
    \hline
    \lstbg{codegray!20}{Suggested Fix} \\
\begin{lstlisting}[
        language=C,
        escapeinside={(*@}{@*)},
        aboveskip=-0.15cm,
        belowskip=-0.3cm
    ]
p = OSSL_PARAM_locate_const(params, OSSL_CIPHER_PARAM_AEAD_TLS1_AAD);
if (p == NULL) {
  /* The parameter is required; if not found, report an error and bail out */
  ERR_raise(ERR_LIB_CIPHER, CIPHER_R_MISSING_AEAD_TLS1_AAD);
  return 0;
}
\end{lstlisting} \\
    \hline
    \lstbg{codegray!20}{Ground Truth Fix} \\
\begin{lstlisting}[
        language=C,
        escapeinside={(*@}{@*)},
        aboveskip=-0.15cm,
        belowskip=-0.3cm
    ]
p = OSSL_PARAM_locate_const(params, OSSL_CIPHER_PARAM_AEAD_MAC_KEY);
...
\end{lstlisting} \\
    \hline
    \end{tabular}

    \vspace{1mm}
    \small
    Because no detailed vulnerability was specified in the prompt, the LLM assumed a NULL‐pointer dereference risk when calling OSSL\_PARAM\_locate\_const. Therefore, it adds a check immediately after this call to prevent any subsequent dereference of a NULL pointer.
\end{table}

\begin{table}[ht]
    \caption{Qwen Fix for CVE-2015-1793 \cite{cve-2015-1793}}
    \label{tab:listing_5}
    \scriptsize
    \setlength{\tabcolsep}{1pt}
    \begin{tabular}{|p{\linewidth-6pt}|}
    \hline
    \hl{Certainly! The provided code snippet contains a potential vulnerability related to memory management, specifically the handling of pointers after freeing memory.} This can lead to undefined behavior, including crashes or security vulnerabilities such as use-after-free errors. Let's break down the original code and then explain the repairs made. ...\\
    \hline
    \lstbg{codegray!20}{Suggested Fix} \\
\begin{lstlisting}[
        language=C,
        escapeinside={(*@}{@*)},
        aboveskip=-0.15cm,
        belowskip=-0.3cm
    ]
if (ok > 0) {
(*@\lstbg{yellow!20}{~~if (xtmp != NULL) \opencb}@*)
(*@\lstbg{yellow!20}{~~~~X509\_free(xtmp);}@*)
(*@\lstbg{yellow!20}{~~~~xtmp = NULL; // Avoid dangling pointer}@*)
(*@\lstbg{yellow!20}{~~\closecb}@*)
  ...
}
\end{lstlisting} \\
    \hline
    \lstbg{codegray!20}{Ground Truth Fix} \\
\begin{lstlisting}[
        language=C,
        escapeinside={(*@}{@*)},
        aboveskip=-0.15cm,
        belowskip=-0.3cm
    ]
if (ok > 0) {
  X509_free(xtmp);
  ...
}
\end{lstlisting} \\
    \hline
    \end{tabular}

    \vspace{1mm}
    \small
    Because no detailed vulnerability was specified in the prompt, the LLM assumed that a use-after-free risk existed in the code and generated a patch to prevent any subsequent dereference of the dangling pointer.
\end{table}

\begin{table}[ht]
    \caption{OpenAI Fix for CVE-2023-48795 \cite{cve-2023-48795}}
    \label{tab:listing_6}
    \scriptsize
    \setlength{\tabcolsep}{1pt}
    \begin{tabular}{|p{\linewidth-6pt}|}
    \hline
    ... The guess is considered wrong if the kex algorithm and/or the host key algorithm are guessed wrong (server and client have different preferred algorithms) or if any of the other algorithms cannot be agreed upon. Otherwise, the guess is considered to be right, and the optimistically \hl{sent packet MUST be handled as the first key exchange packet. However, if the guess was wrong and a packet was optimistically sent by one or both parties, such packets MUST be ignored}, and the appropriate side MUST send the correct initial packet. ... \\
    \hline
    \lstbg{codegray!20}{Suggested Fix} \\
\begin{lstlisting}[
        language=C,
        escapeinside={(*@}{@*)},
        aboveskip=-0.15cm,
        belowskip=-0.3cm
    ]
if (first_kex_packet_follows || session->send_first_kex_follows) {
  ...
  if (session->first_kex_follows_guess_wrong) {
    (*@\lstbgt{red!20}{discard\_first\_kex\_packet(session);}@*)
  } else {
    // Process the optimistically sent packet normally
    (*@\lstbgt{red!20}{process\_first\_kex\_packet(session);}@*)
...
\end{lstlisting} \\
    \hline
    \end{tabular}

    \vspace{1mm}
    \small
    Provided the specification text in the prompt. However, since the code and the specification are at different abstraction levels, the LLM created functions that do not exist in the codebase, resulting in a patch that cannot pass compilation.
\end{table}

\begin{table}[ht]
    \caption{OpenAI Fix for CVE-2019-1543 \cite{cve-2019-1543}}
    \label{tab:listing_7}
    \scriptsize
    \setlength{\tabcolsep}{1pt}
    \begin{tabular}{|p{\linewidth-6pt}|}
    \hline
    \lstbg{codegray!20}{Suggested Fix} \\
\begin{lstlisting}[
        language=C,
        escapeinside={(*@}{@*)},
        aboveskip=-0.15cm,
        belowskip=-0.3cm
    ]
...
(*@\lstbg{yellow!20}{~~if (actx->aad != NULL || actx->len.aad != 0 ||}@*)
(*@\lstbg{yellow!20}{~~~~~~actx->len.text != 0) \opencb}@*)
(*@\lstbg{yellow!20}{~~~~EVPerr(EVP\_F\_CHACHA20\_POLY1305\_CTRL,}@*)
(*@\lstbg{yellow!20}{~~~~~~~~~~~EVP\_R\_COPY\_ERROR);}@*)
(*@\lstbg{yellow!20}{~~~~~return 0;}@*)
(*@\lstbg{yellow!20}{~~\closecb}@*)
  {
    EVP_CIPHER_CTX *dst = (EVP_CIPHER_CTX *)ptr;
    dst->cipher_data = OPENSSL_memdup(
      actx, sizeof(*actx) + Poly1305_ctx_size());
    if (dst->cipher_data == NULL) {
      EVPerr(EVP_F_CHACHA20_POLY1305_CTRL,
             EVP_R_COPY_ERROR);
      return 0;
    }
  }
}
return 1;
\end{lstlisting} \\
    \hline
    \end{tabular}

    \vspace{1mm}
    \small
    The generated patch has different logic from the ground-truth patch. The ground-truth is provided in ~\ref{lst:example1}.
\end{table}

\begin{table}[ht]
    \caption{KNOD \cite{jiang2023knod} Fix for CVE-2019-1543 \cite{cve-2019-1543} (Synthetic)}
    \label{tab:listing_8}
    \scriptsize
    \setlength{\tabcolsep}{1pt}
    \begin{tabular}{|p{\linewidth-6pt}|}
    \hline
    \lstbg{codegray!20}{Suggested Fix} \\
\begin{lstlisting}[
        language=C,
        escapeinside={(*@}{@*)},
        aboveskip=-0.15cm,
        belowskip=-0.3cm
    ]
...
case EVP_CTRL_AEAD_SET_IVLEN:
(*@\lstbg{yellow!20}{~~if (actx == null )\opencb}@*)
(*@\lstbg{yellow!20}{~~~~return 0;}@*)
(*@\lstbg{yellow!20}{~~\closecb}@*)
  if (arg <= 0 || arg > CHACHA_CTR_SIZE) {
    return 0;
  }
  actx.nonce_len = arg;
  return 1;
...
\end{lstlisting} \\
    \hline
    \end{tabular}

    \vspace{1mm}
    \small
    The generated patch has different logic from the ground-truth patch. The ground-truth for the real-world example is provided in~\ref{lst:example1}.
\end{table}

\begin{table}[ht]
    \caption{SimFix \cite{jiang2018shaping} Fix for CVE-2019-1543 \cite{cve-2019-1543} (Synthetic)}
    \label{tab:listing_9}
    \scriptsize
    \setlength{\tabcolsep}{1pt}
    \begin{tabular}{|p{\linewidth-6pt}|}
    \hline
    \lstbg{codegray!20}{Suggested Fix} \\
\begin{lstlisting}[
        language=C,
        escapeinside={(*@}{@*)},
        aboveskip=-0.15cm,
        belowskip=-0.3cm
    ]
...
case EVP_CTRL_AEAD_SET_IVLEN:
  if (arg <= 0 || arg > CHACHA_CTR_SIZE) {
(*@\lstbg{yellow!20}{~~~~if (arg <= 0 || rhType > CHACHA\_CTR\_SIZE) \opencb}@*)
(*@\lstbg{yellow!20}{~~~~~~return 0;}@*)
(*@\lstbg{yellow!20}{~~~~\closecb}@*)
    return 0;
  }
  actx.nonce_len = arg;
  return 1;
...
\end{lstlisting} \\
    \hline
    \end{tabular}
    
    \vspace{1mm}
    \small
    The generated patch has different logic from the ground-truth patch. The ground-truth for the real-world example is provided in~\ref{lst:example1}.
\end{table}


\section{Additional Insights}

\noindent\textbf{Error Analysis of Incorrect Patches.} 
To systematize our error analysis and gain additional insight, we randomly sampled 200 patch instances deemed not plausible (i.e., cases where compilation failed, tests failed, or tests were unavailable) from our off-the-shelf LLM patch-generation experiments. We then manually reviewed each instance to identify the primary reason for failure. From this assessment, we derived four main failure categories: (i) invalid variable/function hallucination, where the patch introduces non-existent variables or function names, usually imagined from the prompt text(typically leading to compilation failures); (ii) incorrect logic, where the patch applies a fundamentally different approach than what is required to fix the vulnerability; (iii) partially correct/incomplete logic, where the patch captures some elements of the intended fix but omits key steps or includes incorrect reasoning; and (iv) presumably correct logic, where the patch appears conceptually correct but cannot be deemed plausible because no test cases are available for validation. The distribution across the 200 assessed patches is shown in Table \ref{tab:root_cause}.

\begin{table}[ht]
\centering
\resizebox{\linewidth}{!}{
\begin{tabular}{|l|c|}
\hline
\textbf{Root Cause Type} & \textbf{Count} \\
\hline
Incorrect Logic & 107 \\
\hline
Partially correct/Incomplete Logic & 57 \\
\hline
Invalid Variable/Function Hallucination & 23 \\
\hline
Presumably Correct Logic & 13 \\
\hline
\end{tabular}
}
\caption{Root Cause Analysis of Incorrect Patches for 200 sampled patch instances.}\label{tab:root_cause}
\end{table}

Of the 200 assessed patches, 112 patches compile successfully. We also observed two qualitative patterns. First, smaller patches leave less room for \emph{partial correctness} missing a single required step often results in a patch being categorized as incorrect logic rather than partially correct. Second, for a given vulnerability instance, multiple patch suggestions often exhibit the same failure mode.

\end{document}